\documentclass[preprint,showpacs,preprintnumbers,amsmath,amssymb,superscriptaddress]{revtex4-2}
\usepackage{graphicx}
\usepackage{amsmath}
\usepackage{amssymb}
\usepackage{epsfig}
\usepackage{amsthm}
\usepackage{color}
\usepackage{bm}

\newtheorem{Theorem}{Theorem}

\def\be{\begin{equation}}
\def\ee{\end{equation}}
\def\arr{\begin{array}{rll}}
\def\ea{\end{array}}
\def\bea{\begin{eqnarray}}
\def\eea{\end{eqnarray}}

\begin{document}
\title{General Three-Body Problem in Conformal Euclidean Space: Hidden Symmetries and Complexity of Low-Dimensional Systems}

\author{Ashot S. Gevorkyan}
\affiliation{Institute for Informatics and Automation Problems NAS of RA,
1, P. Sevak str., Yerevan, 0014, Republic of Armenia}
\affiliation{Institute of Chemical Physics,  NAS of RA, 5/2, P. Sevak str., Yerevan,
 0014, Republic of Armenia}
\author{Aleksander V. Bogdanov}
\affiliation{St. Petersburg State University, 7/9 Universitetskaya nab., St. Petersburg, 199034 Russia}
\author{Vladimir V. Mareev} 
\affiliation{St. Petersburg State University, 7/9 Universitetskaya nab., St. Petersburg, 199034 Russia}
 \date{\today}
\begin{abstract}
Despite the huge number of research into the three-body problem in physics and mathematics, the study of this problem still remains relevant both from the point of view of its broad application and 
taking into account its fundamental significance for the theory of dynamical systems. In addition, to solve the problem of quantum-to-classical transition, it is important to answer the question: is irreversibility 
fundamental to the description of the classical world? To answer this question, we considered a reference classical dynamical system, the general three-body problem, formulating it in conformal Euclidean space 
and rigorously proving its equivalence to the Newtonian three-body problem.  It is shown that a curved configuration space with a local coordinate system reveals new hidden symmetries of the internal motion of 
a dynamical system, which makes it possible to reduce the problem to a 6th order system instead of the known 8th order. The most important consequence of this consideration is that the chronologizing parameter of 
the motion of a system of particles, which we call internal time, is in the general case irreversible, which is characteristic of the general three-body problem. An equation is derived that describes the evolution of 
the flow of geodesic trajectories, with the help of which the entropy of the system is constructed. New criteria for assessing the complexity of a low-dimensional dynamic system and the dimension of stochastic 
fractal structures arising in three-dimensional space are obtained. An effective mathematical algorithm has been developed for the numerical simulation of the general three-body problem, which is traditionally a 
difficult-to-solve system of stiff ordinary differential equations.

\end{abstract}

\maketitle
\section{I\lowercase{ntroduction}}
\label{0}
The initial formulation of the three-body problem in physics and mechanics was to describe the trajectories of particles with point masses by Newton’s equations of motion, taking into account 
the fact that particles interact according to Newton’s law of universal gravitation, when the initial positions and velocities of the bodies are given \cite{Barr}. Over time, it became clear that the 
three-body problem describes a significant number of elementary atomic-molecular processes known to us, which stimulated new research of problem taking into account the multichannel scattering 
of three bodies (see diagram in Fig. 1) and the three-particle interaction between them \cite{Goody}. It is obvious that in this case the computational complexity of the problem increases significantly 
for a number of well-known reasons, especially if we take into account that in many cases elementary processes are irreversible, whereas it is well known that in classical mechanics the time is reversible. 
Recall that this follows from the mathematical formulation of classical mechanics, time $t$ enters through the second-order operator $\partial^2/\partial t^2$, which means that the operator and, 
accordingly, the Equation are invariant under the transformation $t\to-t$. In this regard, the   paper  \cite{Zai}  pose a reasonable question to researchers,  namely: is irreversibility fundamental to 
the description of the classical world?  Note that this question turns out to play a fundamental role in solving the problem of matching the quantum-classical transition \cite{gev,Gev0}.  In other words, 
new research, in addition to the goal of reducing the mathematical complexity of the problem, should answer a number of conceptual questions, and primarily the question of the irreversibility of a 
number of classical elementary atomic-molecular processes associated with time. Ideally, time as a measure of the movement of a dynamic system should accurately reflect the speed, nature and 
singularites of the changes occurring in the system.

Recall that time is one of the basic concepts of philosophy and physics. This is a measure of the duration and continuous existence of all material objects, a characteristic of the sequential change in 
their states in processes and the processes themselves, occurring in an obviously irreversible sequence from the past, through the present to the future  \cite{Oxf,Time,FE}. The twentieth 
century saw two fundamental revolutions in physics - relativity and quantum theory. If in the theory of relativity time was equated in status to coordinates and Minkowski space was introduced 
with four-dimensional invariant interval that determines the distance between two events, then in quantum mechanics it remained within the framework of Newton’s classical definition \cite{New}. 
In any case, it can be said that these new concepts did not significantly change the above classical or Newtonian definition of time.  In particular, relativistic mechanics only deforms and 
compresses time without affecting its essence - one-dimensionality, homogeneity and continuity. 
 \begin{figure}
\includegraphics[width=105mm]{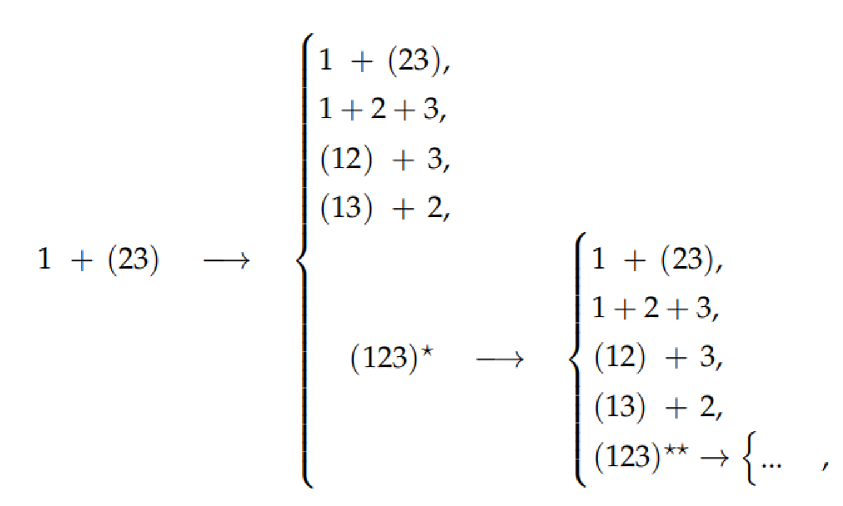}
\caption{ \emph{The problem of multichannel scattering in a classical three-body system can be represented in the most general form, as shown in the diagram, where 1, 2 and 3 denote  
interacting particles,  brackets ($\cdot\cdot\cdot$) denote a coupled system of two bodies, and $(\cdot\cdot\cdot)^\ast$ and $(\cdot\cdot\cdot)^{\ast\ast}$ denote accordingly some 
short-lived coupled three-body system. }}
\label{fig1}
\end{figure}
Unlike the ability of a body to move in opposite directions in space, time is usually considered to have only one dimension and one direction, despite the fact that most laws of physics allow 
that any process can occur in both forward and backward directions. In modern physics, there are only a small number of known physical phenomena that violate the reversibility 
of time, which presupposes the presence of a directionality or arrow of time. Recall that recognized examples of the arrow of time are \cite{Donog,Har,Byrne,Zeh,Hemmo,SpTi,Comp}:
\begin{itemize}
\item{Time arrow, characteristic of expanding waves (such as light or sound emanating from a source).}
\item{The arrow of time associated with the growth of entropy in an isolated system according to the second law of thermodynamics.}
\item{The arrow of time associated with the irreversibility of measurements in quantum mechanics.}
\item{The arrow of time associated with the preference for a certain direction of time in weak interactions in particle physics (CP symmetry breaking)}.
\item{The cosmological arrow of time, which is associated with the expansion of the Universe after the Big Bang.} 
\item{Irreversibility and the arrow of time are characteristic of quantum systems immersed in a fundamental environment, such as a continuously fluctuating quantum vacuum \cite{AshGev,AshGev1}.} 
These systems can be called complex or whole system, which is more than the sum of their parts \cite{Lin}.
\end{itemize}
Note that the fundamentality of the various arrows of time and their interrelationship is a hotly debated topic in modern theoretical physics (see for example \cite{Peter}).
 
Our main goal is to study the issue of irreversibility in classical mechanics and the role of time, or more precisely, the parameter of chronologization of the motion stages of a classical dynamical 
system.  To do this, we considered the general three-body problem, which is quite well studied and in a certain sense is a reference example of a dynamical system. It should also be noted that 
the classical three-body problem plays an important role in the context of quantum-classical matching during the transition of a complex dynamical system from the region of quantum motion to 
the region of classical chaotic motion (Poincar\'{e} systems). The study of this problem is of great practical importance, since a significant number of chemical reactions known and used by us can 
be described within the framework of the three-body model. 

A mathematical feature of the three-body problem is that it is usually described by an eighth-order system with stiff differential equations, the stability of which requires computation with very small 
time-step. The problem becomes even more complicated when we consider a multichannel scattering problem taking into account arbitrary, including three-particle interactions, which is typical for all 
atom-molecular processes. In this case, as a rule, over a wide range of initial data of the problem, the system exhibits chaotic behavior. Note that, due to these difficulties, no good solver has yet 
been created for mass calculations of multichannel atom-molecular processes. 

In other words, new studies of the three-body problem are very relevant both from the point of view of identifying new features of dynamic systems and from the point of view of creating an 
effective package for numerical modeling of the problem.
 
To study this question, we recently considered the  multichannel scattering in a three-body system (see diagram in Fig. 1), formulating it in a conformal Euclidean-Riemanian space  associated with 
the energy surface of the system of interacting bodies \cite{Gev0}. The developed representation made it possible to identify two new hidden symmetries of the internal motion of a three-body system 
and thereby reduce the number of equations of motion to six instead of the usual eight. Note that when moving to a local reference system, a new parameter for chronologizing the phases of motion 
of a dynamical system ``$s$'' (\emph{internal time}) appears in the problem, which is actually the trajectory of the center of mass of a three-body system. Despite the fact that the system of 
fundamental Equations of Motion is invariant under the transformation $s\to -s$ (see below (\ref{q1.01}), in a global or Cartesian coordinate system this chronolization parameter $s\mapsto\mathfrak{s}$ 
has completely different properties from ordinary time ``$t$". It is multidimensional, heterogeneous and oriented, whereas in classical mechanics the time ``$t$'' has an absolute value and does not 
depend on the processes occurring in the system.
 
All this ultimately leads to a more accurate description of the stages of evolution of an irreversible system, in this case the process of  multichannel scattering in a three-body system (see diagram in Fig. 1).  
The article derives an Equation for the nonstationary entropy of the flow of geodesic trajectories in phase space, which differs from the definition of the Kolmogorov-Sinai \cite{Kol1,Sin} and Pesin \cite{Pesin} 
entropy as applied to dynamic systems. Using entropy, a criterion for the complexity of a low-dimensional dynamic system is determined. A new definition of the dimension of the fractal structure of internal 
time,  different from the Hausdorff-Bezikovich criterion, was also studied and given. Our calculations show that internal time $\mathfrak{s}$  immersed in three-dimensional space always has a dimension 
less than  one and is equal to 0.89.

The article is organized as follows:

In Section 2, we present the problem statement and derive differential equations describing the motion of three bodies in a local coordinate system, as well as an underdetermined system 
of algebraic equations that allows us to construct transformations between local and global coordinate systems in differential form. An integral representation for the three-dimensional 
internal time of a three-particle dynamical system in global coordinates Jacobi is constructed.

In Section 3, we analyze in detail the mechanisms of occurrence of randomness in the geodesic trajectories of a deterministic dynamic system and formulate a criterion for the divergence 
of two close geodesic trajectories, similar to the Lyapunov exponent. A system of stochastic differential Equations is determined, with the help of which an second-order partial differential
Equation is obtained that describes the evolution of the distribution of conditional probability in the flow of geodesic trajectories.

In Section 4, we define the nonstationary probability distribution and the normal probability distribution in a geodesic trajectory flow. By analogy with Shannon's entropy, we define the 
entropy of a low-dimensional dynamical system and its measure of complexity. A formula is obtained to calculate the region of classical chaos in phase space. An analogue of the Lyapunov 
exponential for the wave function of a quantum three-body system has been determined.

In Section 5, we present mathematical algorithms for organizing simulation of the problem on high-performance computing machines. A mathematical algorithm for calculating three-
dimensional manifolds is described in detail, using which two complete members of different families of manifolds are calculated and visualized. A mathematical algorithm has been 
created for solving the trajectory problem in a local coordinate system, the potentials of pair interactions between particles have been specified, and the initial conditions for solving 
Cauchy problem have been formulated.

In Section 6, we  discuss the obtained theoretical and numerical results and outline directions for future research.

section{F\lowercase{ormulation of the problem}}

In the 20th century, physics experienced a qualitative transition from the simplicity of reductionism to the complexity of chaos theory, which ultimately fundamentally changed our understanding 
of complex phenomena, ushering in the development of a new discipline - complexity science. Note that pioneering work on chaos theory was carried out by H. Poincar\'{e}  when studying the 
three-body problem of celestial mechanics \cite{Poin}.  He showed that even simple deterministic systems can exhibit aperiodic, chaotic behavior that is extremely sensitive to initial conditions, 
which in turn challenged Laplace's philosophical view of a deterministic universe \cite{Lap}. 

In this work, we consider a more general formulation of the three-body problem, namely multichannel scattering (see diagram in Fig. 1) with arbitrary interactions potential between bodies, including 
taking into account three-particle interactions. In particular, as E. Bruns \cite{Bruns} showed, the problem under consideration is described in an 18-dimensional phase space and has 10 integrals of 
motion. Taking into account these integrals of motion, the number of equations of the dynamical system is reduced to an 8th order system, i.e. up to 8 first order ordinary differential Equations.
Recall an illustrative example from classical physics that demonstrates chaotic behavior due to four coupled first-order ordinary differential equations derived from Hamilton's Equations of motion.
  
As recently shown in the work \cite{Gev0}, the representation developed based on the formulation of the three-body problem on conformal Euclidean space, associated with the energy surface
of the system of bodies, allows us to discover new hidden internal symmetries of the dynamical system. This makes it possible to carry out a more complete integration in the general case of 
the non-integrable three-body problem and reduce it to a 6th order system instead of the usual 8th order.  In the local coordinate system $(x^1,x^2,x^3)=
\{\bar{x}\}\in\mathcal{M}_t^{(3)}$ the dynamical system is described by the following set of first-order stiff \emph{ordinary differential equations} (ODEs):
\begin{equation}
\begin{cases}
\dot{\zeta}^1 =a_1\bigl\{(\zeta^1)^2-(\zeta^2)^2-(\zeta^3)^2-\Lambda^2\bigr\} +2\zeta^1\bigl\{a_2\zeta^2+a_3\zeta^3\bigr\}, \qquad \zeta^1=\dot{x}^1,
\\
\dot{\zeta}^2 =a_2\bigl\{(\zeta^2)^2-(\zeta^3)^2-(\zeta^1)^2-\Lambda^2\bigr\} +2\zeta^2\bigl\{a_3\zeta^3+a_1\zeta^1\bigr\}, \qquad \zeta^2=\dot{x}^2, 
\\
\dot{\zeta}^3=a_3\bigl\{(\zeta^3)^2-(\zeta^1)^2-(\zeta^2)^2-\Lambda^2\bigr\} +2\zeta^3\bigl\{a_1\zeta^1+a_2\zeta^2\bigr\}, \qquad \zeta^3=\dot{x}^3,
\label{q1.01}
\end{cases}
\end{equation} 
where $\mathcal{M}^{(3)}_t$ is a bundle of the three-dimensional manifold $\mathcal{M}^{(3)}$, which is defined by the metric tensor $g_{ij}(\{\bar{x}\})=\delta_{ij}[E-U(\{\bar{x}\})]/U_0>0,\,\, (i,j=1,2,3$, 
in addition, $E$ and $U(\{\bar{x}\})$ are the total energy and potential energy of the system of bodies, respectively, $U_0=\max U(\{\bar{x}\})=const$, and $\delta_{ij}$ is the Kronecker delta). In the system 
of equations (\ref{q1.01})  $\dot{\zeta}^i=d\zeta^i/ds$ denotes the derivative with respect to the chronolizing parameter ``$s$'' or time in the system $\{\bar{x} \}$, which is determined using the sum of integrals:
\begin{equation}
 s(\{\bar{x}\})=\sum_{i=1}^3\int\sqrt{g(\{\bar{x}\})} dx^i.
 \label{q2.02a}
\end{equation} 
In addition, the following notations are made:
\begin{equation}
 a_i(\{\bar{x}\})=-\frac{\partial\ln\sqrt{g(\{\bar{x}\})}}{\partial{x^i}},\quad \Lambda(\{\bar{x}\})=\frac{J}{g(\{\bar{x}\})}, \quad g(\{\bar{x}\})=\frac{[E-U(\{\bar{x}\})]}{U_0}>0.
\label{q2.02}
\end{equation} 
where $U(\{\bar{x}\})$ and $E$ respectively denote the resulting interaction potential between bodies and the total energy of the system,  $U_0=\max | U(\{\bar{x}\})|=const$ and $J=const$ 
is the total angular momentum  of three-body system.

It is easy to check that the system of Equations (\ref{q1.01}) is formally invariant under the change $s\to-s$, i.e. is reversible with respect to the chronolizing parameter -internal time in the 
local reference frame $s(\{\bar{x}\})$. Note that $s(\{\bar{x}\})$ being a function of local coordinates and during the evolution of a dynamical system represents a continuous and 
one-dimensional curve, which in many features is equivalent to the definition of the usual time, i.e. $s(\{\bar{x}\})\equiv t$.
\begin{figure}
\includegraphics[width=75mm]{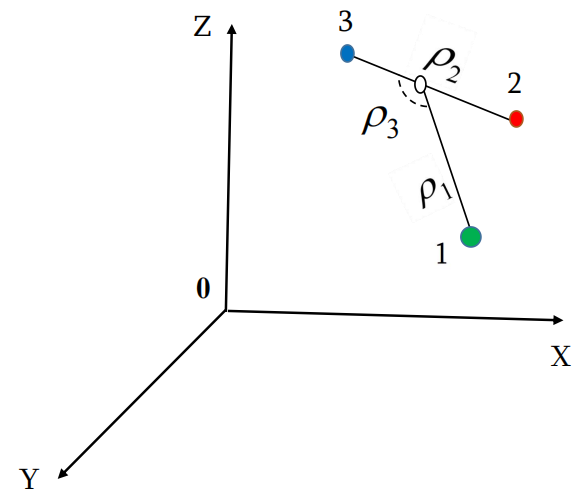}
\caption{\emph{In the Cartesian coordinate system $(X,Y,Z)$ the Jacobi coordinates $\{\bar{\rho}\}=(\rho_1,\rho_2,\rho_3)$ are shown, where the colored circles indicate bodies 1, 2 and 3, and 
the colorless circle respectively indicates the center of mass of bodies 2 and 3.}}
\label{Fig2}
\end{figure}
For a more detailed and understandable study of the  properties of internal time  we need to move to the global coordinate system or the Jacobi coordinate system
$\{\bar{\rho}\}=(\rho_1,\rho_2, \rho_3)\in\mathbb{R}^3$, where $\mathbb{R}^3$ is three-dimensional Euclidean space (see Fig. 2). In this system, one can reproduce the explicit 
form of the metric tensor element $g(\{\bar{x}\})=\overline{g}(\{\bar{\rho}\})$ and get  the opportunity to comprehensively study the structure and the nature of the chronological 
parameter $s(\{\bar{x}\})\mapsto\mathfrak{s}(\{\bar{\rho}\})$. 

A completely different situation arises when the chronologizing parameter is considered in a global coordinate system $\mathfrak{s}(\{\bar{\rho}\})$. During the evolution of the dynamical 
system, this parameter, as a continuous curve, will fill the space $\mathbb{R}^3$, in a sense similar to the three-dimensional Hilbert curve \cite{Hilbert}. Moreover, its two-dimensional 
projections can have non-trivial topological features characterized by Betti number \cite{Spanier}, and metric properties will be described by invariant measures and fractal dimensions \cite{Badii}. 
In what follows, the chronologizing parameter in the global system $\mathfrak{s}(\{\bar{\rho}\})$  will be called \emph{internal time}  of the three-particle dynamical system.

As shown, transformations between of local $\{\bar{x}\}$ and global $\{\bar{\rho}\}$ coordinate systems can only be represented in differential form \cite{Gev0}:
\begin{equation}
\begin{cases}
d\rho_1=\alpha_1dx^1+\alpha_2dx^2+\alpha_3dx^3,
\\
d\rho_2=\beta_1dx^1+\beta_2dx^2+\beta_3dx^3,
\\
d\rho_3=\lambda_1dx^1+\lambda_2dx^2+\lambda_3dx^3,
\label{q1.03}
\end{cases}
\end{equation} 
where the coefficients $(\alpha_1,\cdot\cdot\cdot,\beta_1,\cdot\cdot\cdot,\lambda_3)$ are solutions to an underdetermined system of algebraic Equations of the form:
\begin{eqnarray}
\alpha^2_1+\beta_1^2+\lambda_1^2=\overline{g}(\{\bar{\rho}\}),\qquad \alpha_1\alpha_2+\beta_1\beta_2+\lambda_1\lambda_2=0,
\nonumber\\
\alpha^2_2+\beta_2^2+\lambda_2^2=\overline{g}(\{\bar{\rho}\}),\qquad  \alpha_1\alpha_3+\beta_1\beta_3+\lambda_1\lambda_3=0,
\nonumber\\
\alpha^2_3+\beta_3^2+\lambda_3^2=\overline{g}(\{\bar{\rho}\}),\qquad   \alpha_2\alpha_3+\beta_2\beta_3+\lambda_2\lambda_3=0.
\label{q1.04}
\end{eqnarray} 
It is not difficult to verify that the system of underdetermined algebraic Equations (\ref{q1.04}), which consists of 6 Equations and 9 unknowns, can generate three-dimensional  oriented smooth 
manifolds  $\mathcal{R}^{(3)}$, which are immersed in nine-dimensional Euclidean space $\mathbb{R}^9$. The number of such manifolds can be easily calculated using the well-known combinatorial
formula $C^m_n=\frac{n!}{m!(n-m)!}$, which determines the number of $n$-permutations of the set $ m$.  Considering that $n=9$ denotes the number of unknowns and $m=6$, respectively, 
the number of Equations, the number of generated  manifolds will be 84. As shows from the analysis of the symmetry of the Equations  system  (\ref{q1.04}), only four families of manifolds 
are possible $\eta=(\mathcal{A},\mathcal{B},\mathcal{C},\mathcal{D})$. For definiteness, below we will consider the first two families, each of which consists of six manifolds;
$ 
\mathcal{A}=(\mathcal{A}_1,\mathcal{A}_2,\mathcal{A}_3,\mathcal{A}_4,\mathcal{A}_5,\mathcal{A}_6)
$ 
and
$\mathcal{B}=(\mathcal{B}_1,\mathcal{B}_2,\mathcal{B}_3,\mathcal{B}_4,\mathcal{B}_5,\mathcal{B}_6)$, where each family member is determined by three independent parameters, such as:
$$ 
\mathcal{A}=\bigl\{(\alpha_1,\alpha_2,\alpha_3),(\beta_1,\beta_2,\beta_3),(\lambda_1,\lambda_2,\lambda_3),(\alpha_1,\beta_1,\lambda_1),(\alpha_2,\beta_2,\lambda_2),(\alpha_3,\beta_3,\lambda_3)\bigr\},
$$ 
and respectively for the family:
$$ 
\mathcal{B}=\bigl\{(\alpha_1,\beta_2,\lambda_3),(\alpha_1,\beta_3,\lambda_2),(\alpha_2,\beta_1,\lambda_3),(\alpha_2,\beta_3,\lambda_1),(\alpha_3,\beta_1,\lambda_2),(\alpha_3,\beta_2,\lambda_1)\bigr\}.
$$ 
Finally we can combine each family of manifolds as a direct sum of sets:
$$
\mathcal{R}^{(3)}_\mathcal{A}\cong\bigcup_{i=1}^6 \mathcal{R}^{(3)}_{\mathcal{A}_i},  \qquad \mathcal{R}^{(3)}_\mathcal{B}\cong\bigcup_{i=1}^6 \mathcal{R}^{(3)}_{\mathcal{B}_i}.
$$
In a similar way, we can construct different families of manifolds. 

Finally, it is important to note that each manifold of a given family is also surrounded by two additional manifolds, the direct products of which form a nine-dimensional manifold. In particular, the manifold 
$ \mathcal{R}^{(3)}_{\mathcal{A}_1}=\mathcal{R}^{(3)}_{(\alpha_1,\alpha_2,\alpha_3)}$ is 
surrounded by two additional manifolds $ \mathcal{R}^{(3)}_{(\beta_1,\lambda_2,\beta_3)}$ and $ \mathcal{R}^{(3)}_{(\lambda_1,\beta_2,\lambda_3)}$,  which can be combined using the direct product 
$ \mathcal{R}^{(9)}_{\mathcal{A}_1}= \mathcal{R}^{(3)}_{\mathcal{A}_1}\times \mathcal{R}^{(3)}_{(\beta_1,\lambda_2,\beta_3)}\times \mathcal{R}^{(3)}_{(\lambda_1,\beta_2,\lambda_3)}$.
Below we will call $ \mathcal{R}^{(9)}_{\mathcal{A}_1}$ a complete member of the $\mathcal{A}$ family. As for the complete member of the family $\mathcal{B}$, it can be represented in the form
 $ \mathcal{R}^{(9)}_{\mathcal{B}_1}= \mathcal{R}^{(3)}_{\mathcal{B}_1}\times \mathcal{R}^{(3)}_{(\alpha_2,\beta_3,\lambda_1)}\times \mathcal{R}^{(3)}_{(\alpha_3,\beta_1,\lambda_2)}$,
where $\mathcal{R}^{(3)}_{\mathcal{B}_1}=\mathcal{R}^{(3)}_{(\alpha_1,\beta_2,\lambda_3)}$.

Now let us move on to constructing the form of internal time in the global coordinate system $\mathfrak{s}(\{\bar{\rho}\})$. Using the inverse transformation, we can express increments in 
local coordinates in terms of increments in global Jacobi coordinates:
\begin{equation}
\begin{cases}
dx^1=\breve{\alpha}_1d\rho^1+\breve{\alpha}_2d\rho_2+\breve{\alpha}_3d\rho_3,
\\
dx^2=\breve{\beta}_1d\rho^1+\breve{\beta}_2d\rho_2+\breve{\beta}_3d\rho_3,
\\
dx^3=\breve{\lambda}_1d\rho^1+\breve{\lambda}_2d\rho_2+\breve{\lambda}_3d\rho_3,
\label{q1.05}
\end{cases}
\end{equation} 
where the coefficients $(\breve{\alpha}_1,\cdot\cdot\cdot,\breve{\alpha}_3,\cdot\cdot\cdot,\breve{\lambda}_3)$ are determined by the following formulas:
\begin{eqnarray}
\breve{\alpha}_1=(\beta_2\lambda_3-\beta_3\lambda_2)A,\qquad \breve{\alpha}_2=(\beta_3\lambda_1-\beta_1\lambda_3)A,\qquad \breve{\alpha}_3=(\beta_1\lambda_2-\beta_2\lambda_1)A,
\nonumber\\
\breve{\beta}_1=(\alpha_3\lambda_2-\alpha_2\lambda_3)A,\qquad \breve{\beta}_2=(\alpha_1\lambda_3-\alpha_3\lambda_1)A,\qquad \breve{\beta}_3=(\alpha_2\lambda_1-\alpha_1\lambda_2)A,
\nonumber\\
\breve{\lambda}_1=(\alpha_2\beta_3-\alpha_3\beta_2)A,\,\qquad \breve{\lambda}_2=(\alpha_3\beta_1-\alpha_1\beta_3)A,\qquad \breve{\lambda}_3=(\alpha_1\beta_2-\alpha_2\beta_1)A,
\label{q1.06}
\end{eqnarray} 
in addition, $A=1/[\alpha_1\beta_2\lambda_3+\alpha_2\beta_3\lambda_1+\alpha_3\beta_1\lambda_2-\alpha_1\beta_3\lambda_2-\alpha_2\beta_1\lambda_3-\alpha_3\beta_2\lambda_1]$ is a determinant.
Now, taking into account the equation (\ref{q2.02a}), for internal time we get the following expression:
\begin{eqnarray} 
\mathfrak{s}(\{\bar{\rho}\})={s}_0+{\alpha}\int_{{\rho}^0_1}^{{\rho}_1}\sqrt{\bar{g}(\{\bar{\rho}\})}d\rho_1+{\beta}\int_{{\rho}^0_2}^{{\rho}_2}\sqrt{\bar{g}(\{\bar{\rho}\})}d\rho_2
+{\lambda}\int_{{\rho}^0_3}^{{\rho}_3}\sqrt{\bar{g}(\{\bar{\rho}\})}d\rho_3,
\label{q1.07}
\end{eqnarray}
 where the following notations are made; ${\alpha}=\sum_{i=1}^3\breve{\alpha}_i$,\, ${\beta}=\sum_{i=1}^3\breve{\beta}_i$ and ${\lambda}=\sum_{i=1}^3\breve{\lambda}_i$.

Note that the internal time $\mathfrak{s}(\{\bar{\rho}\})$ in the form of a curve in three-dimensional space appears in the asymptotic subspace $(in)$, then goes into one of the asymptotic subspace 
$( out ) $, while it is wound in a complicated way around of the selected curve $\mathfrak{s}_i,\,i=\overline{1,3}$ connecting two asymptotic subspaces (see Fig. 3).
 \begin{figure}
\includegraphics[width=100mm]{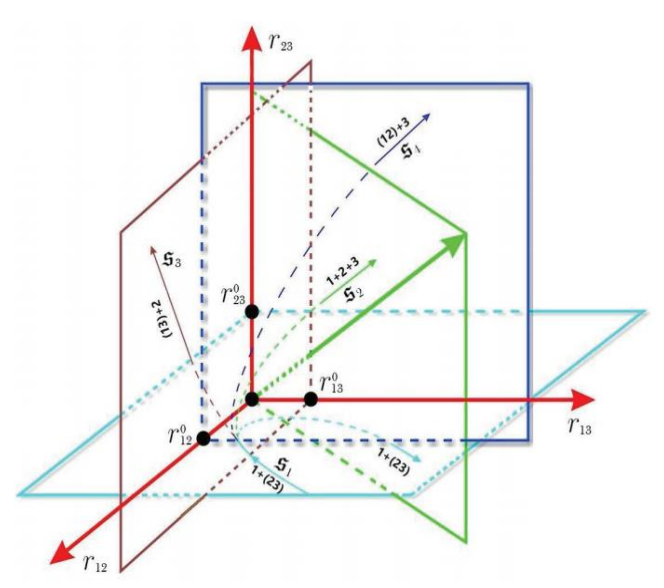}
\caption{\emph{The set of smooth curves $\mathfrak{s}=(\mathfrak{s}_1,\,\mathfrak{s}_2,\,\mathfrak{s}_3,\,\mathfrak{s}_4)$ connecting the asymptotic subspace $(in)$, in which the three-body system 
$1+(23)$ is grouped, with other asymptotic subspaces $(out)$, where the particles are grouped as follows:
 $1 + (23)$, \,$(12)+3$, \,$(13)+2$, and $1+2+3$.  The distance between particles ``$i$'' and ``$j$'' in the Cartesian coordinate 
system is given by the expression $r_{ij}(\{\bar{\rho}\})$\,\,$(i,j = 1,2, 3, \, i\neq{j})$, and $r^0_{ij}$ - the average distance between particles in the corresponding pairs. During the scattering process, the 
three-dimensional internal time $\mathfrak{s}(\{\bar{\rho}\})$, which has an arrow, selects a specific asymptotic subspace for transition, which in some conditions may be random.}}
\label{Fig3} 
\end{figure}   
 
Thus, we have formulated all the necessary mathematical formulas for studying the general three-body problem (see Fig. 1) by calculating the trajectory of the reduced mass of three bodies 
$\mu_0=\sqrt{\frac{m_1m_2m_3}{m_1+m_2+m_3}} $, where $m_1,\, m_2$ and $m_3$ in three-dimensional configuration space $\mathbb{R}^3$.

\section{Flow of geodesic trajectories in six-dimensional phase space}

\subsection{Lyapunov exponent for geodesic trajectories flow}
As the analysis shows, a multichannel process (see Fig. 1) in a three-body system is not only irreversible in the general case, but can also exhibit random behavior during its evolution. To study the nature of 
motion in a three-particle dynamical system, one can, like the Lyapunov exponent \cite{Ben}, introduce a parameter showing the degree of divergence of close geodesic trajectories. In particular, the average 
growth rate of the distance between neighboring trajectories of the center of mass of  bodies system is determined by the leading Lyapunov exponent, which for a long (but not too long) time $t$ can be 
estimated as:
\begin{equation} 
\epsilon(t)\simeq \frac{1}{t}\cdot\ln\Bigl\Vert\frac{\Delta\mathfrak{s}(t)}{\Delta\mathfrak{s}(0)}\Bigr\Vert, 
\label{q2.G07}
\end{equation}
where $s\equiv t$ is one-dimensional and homogeneous time, and $\Delta\mathfrak{s}( t) =\mathfrak {s}(t)-\mathfrak{s}'(t)$ denotes the difference of two close internal times or geodetic trajectories,  in addition, 
$\Delta\mathfrak{s}(0)\ll1$ is the distance between two trajectories at the initial moment $t=0$. In what follows, $\epsilon(t)$ will be called the leading time-dependent Lyapunov  exponent. Note that the time-dependent 
Lyapunov exponent, as we will see below, will play a key role in studying the properties of the geodesic trajectories flow in phase space.

{\bf Definition 1}. \emph{Let us call the standard leading Lyapunov exponent the parameter characterizing the asymptotic behavior of the sequence $\bigl\Vert\frac{\Delta\mathfrak{s}(t)}{\Delta\mathfrak{s}(0)}\bigr\Vert$}:
\begin{equation} 
{\epsilon}:=\overline{\lim_{t\to\infty}}\biggl[\frac{1}{t}\cdot\ln\Bigl\Vert\frac{\Delta\mathfrak{s}(t)}{\Delta\mathfrak{s}(0)}\Bigr\Vert\biggr],\qquad t\neq0,
\label{q2.07}
\end{equation}
\emph{where  $a:=b$ means that $a$ is assumed to be equal to $b$ and  the constant $\epsilon=const$, which can be either positive or negative and equal to $0$. }

\subsection{Equation of motion of geodesic trajectories flow in phase space}

Since the system of equations (\ref{q1.01}) is rigid and, moreover, nonlinear, it is very difficult to expect that to solve it it will always be possible to create a more or less reliable mathematical algorithm that will ensure the 
stability of numerical calculations.  In the case when the time-dependent Lyapunov exponent is positive, that is $\epsilon(t)>0$ (see (\ref{q2.G07})), the system experiences chaotic motion, so it makes sense not to study the 
problem of single geodesic trajectories, but to consider the evolution of the flow of geodesic trajectories on the corresponding three-dimensional manifolds. 

In other words, it makes sense to consider a trajectory problem with a random external influence, which for the three-body problem can be written as a system of stochastic differential Equations of the Langevin type
(see also Equations of motion in conformal Euclidean space (\ref{q1.01})):
\begin{equation} 
 \dot{z}^\mu=A^\mu(\{\bar{x}\},\{\dot{\bar{x}}\}) +  f^\mu(t),\qquad \mu=\overline{1,6},
\label{q2.0t0}
\end{equation}
where the following notations are made for the independent variables:
 $$\{z\}=\bigl(z^1=\dot{x}^1;\quad z^2=\dot{x}^2;\quad z^3=\dot{x}^3;\quad z^4={x}^1;\quad z^5={x}^2;\quad z^6={x}^3;\bigr)$$
and for the terms $A^\mu(\{\bar{x}\},\{\dot{\bar{x}}\})$ included in the equation (\ref{q2.0t0}), respectively, the following notations:
\begin{equation}
\begin{cases}
A^1 =a_1\bigl\{(\dot{x}^1)^2-(\dot{x}^2)^2-(\dot{x}^3)^2-\Lambda^2\bigr\} +2\dot{x}^1\bigl\{a_2\dot{x}^2+a_3\dot{x}^3\bigr\}, \qquad A^4=x^1,
\\
A^2 =a_2\bigl\{(\dot{x}^2)^2-(\dot{x}^3)^2-(\dot{x}^1)^2-\Lambda^2\bigr\} +2\dot{x}^2\bigl\{a_3\dot{x}^3+a_1\dot{x}^1\bigr\}, \qquad A^5=x^2, 
\\
A^3=a_3\bigl\{(\dot{x}^3)^2-(\dot{x}^1)^2-(\dot{x}^2)^2-\Lambda^2\bigr\} +2\dot{x}^3\bigl\{a_1\dot{x}^1+a_2\dot{x}^2\bigr\}, \qquad A^6=x^3.
\label{q2t.01}
\end{cases}
\end{equation}
For definiteness, below we will use a random Gauss-Markovian process as the forces $f^\mu(t)$:
\begin{equation} 
\langle f^\mu(t)\rangle=0,\qquad \langle f^\mu(t)f^\mu(t')\rangle=\epsilon(t) \,\delta(t-t'),
\label{q2.0t7}
\end{equation}
where $\epsilon(t)>0$ is the degree of the random process, which in the problem under consideration is a time-dependent Lyapunov exponent (see Equation (\ref{q2.G07})).

Thus, in the deterministic three-body problem, randomities  have been identified that arise due to the influence of the nonlinearity of the equations (\ref{q1.01})) and the underdetermination of 
the system of algebraic equations (\ref{q1.04}), providing transitions between the local and global coordinate systems. To carry out analytical studies, we assumed that the emerging randomness 
$f^\mu(t)$ obeys the Gauss-Markovian process and are generated from a single source.

To exclude the occurrence of other unnatural random influences on the dynamical system, the equations (\ref{q2.0t0})-(\ref{q2.0t7})  must be defined in a local coordinate system along regular 
curves on the manifolds $\mathcal{R}^{(9)}_{\mathcal{A}_1}$ and $\mathcal{R}^{(9)}_{\mathcal{B}_1}$, respectively.

\begin{Theorem} If we assume that geodesic trajectories in phase space are described by the equations (\ref{q2.0t0})-(\ref{q2t.01}), and random forces satisfy the correlation 
conditions (\ref{q2.0t7}), then the geodesic flow in phase space will satisfy the following Fokker-Planck type equation:
\begin{equation} 
\frac{\partial \mathcal{P}}{\partial t}= \sum^6_{\mu=1}\biggl\{\epsilon(t)\frac{\partial^2 }{\partial z^2_\mu}+\frac{\partial }{\partial z_\mu}A^\mu(\{z\})\biggr\}\mathcal{P}.
\label{q2.0t9}
\end{equation}

\begin{proof}
 Let us consider the following functional that describes the evolution of the conditional probability distribution of a geodesic flow in phase space:
\begin{equation} 
\mathcal{P}(\{z\},t;\{z'\},t')= \Bigl\langle\prod^6_{\mu=1}\delta[z^\mu(t)-z^\mu(t')] \Bigr\rangle.
\label{q2.0t10}
\end{equation}
Differentiating expression (\ref{q2.0t10}) with respect to the usual time ``$t$”, taking into account Equation (\ref{q2.0t0})-(\ref{q2t.01}), we obtain:
\begin{eqnarray} 
\partial_t\mathcal{P}(\{z\},t;\{z'\},t')=-\sum^6_{\nu=1}\partial_{z^\nu} \Bigl\langle z^\nu_t\prod^6_{\mu=1}\delta[z^\mu(t)-z^\mu(t')]\Bigr\rangle=
\nonumber\\
\sum^6_{\nu=1}\partial_{z^\nu}\Bigl\{{\bf A}(\{z\},\{\dot{z}\})\mathcal{P}(\{z\},t;\{z'\},t')+\bigl\langle f(t)\delta[z^\mu(t)-z^\mu(t')]\Bigr\},
\label{q2.0t11}
\end{eqnarray}
where
\begin{equation}
{\bf A}(\{z\},\{\dot{z}\})=
\begin{cases}
A^1(\{z\},\{\dot{z}\}),
\\
\quad\cdot\,\,\,\, \cdot\,\,\,\,\cdot\quad\,\,\,,
\\
A^6(\{z\},\{\dot{z}\}).
\label{q2.z01}
\end{cases}
\end{equation} 
Taking into account that the  random process $f(t)$ satisfies the correlation relations (\ref{q2.0t7}), we can calculate the second term in expression (\ref{q2.0t11}).  In particular, using Wick's theorem for an arbitrary functional
$N(\{z\},f(t))$, one can obtain:
\begin{equation}
\bigl\langle f(t)N(\{z\},t;\{z'\},t')\bigr\rangle =2\biggl\langle\frac{\delta N}{\delta f(t)}\biggr\rangle=2\frac{\partial\,\,}{\partial {z^\mu}}\biggl\langle \frac{\delta z^\mu(t)}{\delta f(t)}\delta(z^\mu(t)-z^\mu(t')\biggr\rangle.
\label{q3.01}
\end{equation} 
Recall that $\{z\}=(z^1,\cdot\cdot\cdot,z^6)$ denotes a set of stochastic functions whose variational derivatives with respect to the independent random force $f(t)$ are defined as follows:
\begin{equation}
\biggl\langle\frac{\delta z^\mu(t)}{\delta f(t)}\biggr\rangle=\epsilon(t) \cdot \mathsf{sgn}(t-t')+O(t-t').
\label{q3.0z1}
\end{equation} 
After carrying out the regularization procedure in the sense of the Fourier expansion, we find its value at the time: 
\begin{equation}
t=t':\mathsf{sgn}(0)=\frac{1}{2}.
\label{q3.0zk1}
\end{equation} 
Taking into account the Equalities in (\ref{q2.0t11})-(\ref{q3.0zk1}) for the conditional probability, we obtain the Fokker–Planck Equation of the form (\ref{q2.0t9}). 

Note that the Equation for the evolution of the probability density of the geodesic flow (\ref{q2.0t9}) in phase space is given in local coordinates $\{\bar x\}$ in which the chronolization parameter ``$t$'' changes as ordinary, 
one-dimensional and homogeneous time along the numerical axis from the past to the future through the present. Using transformations of local coordinates (\ref{q1.03}), it is possible to set the interaction potential 
between particles in global Jacobi coordinates $\{\bar{\rho}\}$ and calculate the evolution of the geodesic trajectories flow in phase space, including during the transition of a dynamical system between two asymptotic subspaces. 
\end{proof}
\end{Theorem}

\section{Entropy of the flow of geodesic trajectories and complexity criterion }
Shannon entropy is an effective dynamical indicator that provides a direct measure of the rate of diffusion of geodesic flow, hence the time scale of instabilities encountered when dealing with chaos.
Note that this allows us to more reliably explain the nature of a nonlinear dynamical system characterized by complexity and nonequilibrium.

{\bf Definition 2}.  \emph{Let us call the probability density of the geodesic flow in phase space the expression:}
\begin{equation}
\mathcal{P}^0(\{z\},t)=\mathcal{P}(\{z\},t;\{z'\},t')\bigl|_{\{z'\}=0,\,t'=0},
\label{q3.0zt}
\end{equation} 
\emph{which  is normalized to unity.}

{\bf Definition 3}. \emph{Let us call the equilibrium probability distribution of geodesic flow in phase space the expression:
\begin{equation}
\bar{\mathcal{P}}^0(\{z\},t)=\mathcal{P}(\{z\},t;\{z'\},t')\bigl|_{[\epsilon=\epsilon_0,\,\{z'\}=0,\,t'=0]},
\label{q3.N0zt}
\end{equation}
where $\epsilon_0=\int_0^T\epsilon(t)dt\bigr/T$, in addition, $T$ denotes some finite time interval. }

By analogy with Shannon entropy \cite{Shan}, which characterizes the average rate of information creation by a stochastic data source, for a continuous flow of geodesic trajectories of a low-dimensional 
dynamical system, entropy can be introduced in the form:
\begin{equation}
\mathcal{S}(t)=-\int_{\mathbb{R}^6} \mathcal{P}^0(\{z\},t)\ln\mathcal{P}^0(\{z\},t)dz^1{\cdot\cdot\cdot}dz^6,
\label{q3.K0wt}
\end{equation}
where $\mathbb{R}^6\ni \overline{z^1,z^6}$ denotes the phase space.

\subsection{The three-body problem as a low-dimensional  system with complexity }
The complexity of systems is usually associated with the difficulty of understanding and describing them. However, after Kolmogorov formulated the concept of computational complexity \cite{Kol}, it became 
clear that the complexity of a system could be associated with information \cite{Haken}. Subsequently, complexity also became associated with open systems and with the unpredictable behavior of highly 
nonlinear systems \cite{Morin,Nicolis}. Recently a new meaning has emerged regarding systems theory \cite{vonBert}, which states that complexity is half the way to ``equilibrium'' and ``disequilibrium'' \cite{Kaneko}. 
As we noted above, in a  three-particle system, under certain conditions, non-integrability and chaos can arise. In this case, it is convenient to describe the dynamical system in phase space using probabilistic 
methods (see  (\ref{q2.0t9})).

Since a three-particle dynamical system under certain conditions can have probabilistic properties, then for it, by analogy with a statistical ensemble, a criterion or measure of complexity can be introduced:
\begin{equation}
\mathcal{C}(t)=\mathcal{S}(t)\mathcal{K}(t),
\label{q3.N0wt}
\end{equation}
where $\mathcal{S}(t)$ is  ``information'', which is defined by the Equation (\ref{q3.K0wt}) and a shift from the quilibrium probability distribution (\ref{q3.N0zt}) or  ``disequilibrium" $\mathcal{K}(t)$, which 
is defined as follows:
\begin{equation}
\mathcal{K}^{1/2}(t)=\Bigl|\int_{\mathbb{R}^6} \Delta\mathcal{P}^0(\{z\},t)dz^1{\cdot\cdot\cdot}dz^6\Bigr|,
\label{q5.N0wt}
\end{equation}
where $\Delta\mathcal{P}^0(\{z\},t)=\mathcal{P}^0(\{z\},t)-\bar{\mathcal{P}}^0(\{z\},t).$

Fundamentally important information about the behavior of the quantum three-body problem can be obtained by integrating the probability distribution of the geodesic flow over phase space:
\begin{equation}
{I}^0(t)=\int_{\mathbb{R}^6}\breve{{\mathcal{P}}}^0(\{z\},t)dz^1{\cdot\cdot\cdot}dz^6,
\label{q6.N0wt}
\end{equation}
where $\breve{{\mathcal{P}}}^0(\{z\},t)$ is the probability distribution normalized to unity at the initial time $t=0.$

Recall that the integral (\ref{q6.N0wt}) determines the average volume of the phase space subject to classical chaos. In the case, when the classical region of chaos in the phase space exceeds 
the size of the quantum cell $\hbar^n$ (where $2n$ is the dimension of the phase space), which in this case is equivalent to the inequality $I^0(t)>\hbar^3$, then quantum averaging does not 
eliminate chaos in the quantum system (see \cite{Shust}). Moreover, in this case, the main object of quantum mechanics—the wave function of the system—becomes chaotic. It is obvious that in this case 
the quantization of the classical dynamical system must be carried out differently from the standard approach \cite{Gev0}.

In this case, for low-dimensional quantum chaotic systems, a criterion similar to the Lyapunov exponent can be determined. In particular, for finite time intervals this exponent can be determined by the following formula:
\begin{equation} 
{\epsilon}_q(t)\simeq  \frac{1}{t}\cdot\ln\biggl\Vert\frac{\overline{\delta{\Psi}(\mathfrak{s})}}{\overline{\delta{\Psi}(0)}}\biggr\Vert,\qquad \overline{\delta{\Psi}}(\mathfrak{s})=\int_{\mathcal{M}^{(3)}}\Delta{\Psi}(\mathfrak{s})dz^4dz^5dz^6,\quad t\neq0,
\label{q7G.07}
\end{equation}
while in the limit of large times the exponent will tend to a finite limit:
\begin{equation} 
{\epsilon}_q:=\overline{\lim_{t\to\infty}}\biggl[\frac{1}{t}\cdot\ln\biggl\Vert\frac{\overline{\delta{\Psi}(\mathfrak{s})}}{\overline{\delta{\Psi}(0)}}\biggr\Vert\biggr], \qquad t\neq0,
\label{q7.07}
\end{equation}
where $\Psi(\mathfrak{s},\{\bar{z}\})$ is the wave function of a three-body system, and $\Delta{\Psi}(\mathfrak{s})=[\Psi(\mathfrak{s},\{\bar{z}\})-\Psi(\mathfrak{s}',\{\bar{z}\})]$ denotes the 
difference between two wave functions that were close at the initial time $\mathfrak{s}=0$ and $\{\bar{z}\}=(z^4,z^5,z^6)$.

\section{Mathematical algorithm for numerical simulation of the three-body problem}

\subsection{Model of pair interaction potentials between particles}
As we see, the system of ordinary differential equations (\ref{q1.01}) can be simulated numerically after specifying the interaction potential between bodies in the global coordinate system.
In particular, for definiteness we will assume that the interaction between two arbitrary pairs of bodies is carried out by the Morse potential:
\begin{equation}
U(r_{ij})=U^{(0)}_{ij}\bigl[1-\exp\bigl\{-b_{ij}(r_{ij}-r^0_{ij})\bigr\}\bigr]^2,\qquad i,j=1,2,3,\quad i\neq j,
\label{qV.01}
\end{equation}
where $r_{ij}$ is the distance between particles ``$i$” and  ``$j$”,  the term $U^{(0)}_{ij}=const$ denotes the depth of the interaction potential, $b_{ ij}>0 $ is a certain constant specific to a particular pair 
of interacting particles and $r^0_{ij}$ is the equilibrium communication distance. Recall that the Morse potential is often used in molecular physics because it well describes the interaction of two atoms in 
a bound state within a diatomic molecule.

 Note that these potentials must be written in Jacobi coordinate system, which can be done using the following coordinate transformations:
\begin{eqnarray}
r_{12}=\sqrt{\rho_1^2+(\mu_-\rho_2)^2-2\mu_-\rho_1\rho_2\cos\vartheta},\qquad\qquad\qquad
\nonumber\\
 r_{13}=\sqrt{\rho_1^2+(\mu_+\rho_2)^2+2\mu_+\rho_1\rho_2\cos\vartheta},\qquad\, r_{23}=\rho_2,
\label{qV.02}
\end{eqnarray}
where $\mu_-=m_3/(m_2+m_3)$ and $\mu_+=m_2/(m_2+m_3)$, in addition, $m_2$ and $m_3$  are the particle masses.

To carry out specific calculations it is necessary to determine the full potential of interaction between particles $\overline{g}(\{\bar{\rho}\})=[E-U_{12}(\{\bar{\rho})-U_{13}(\{\bar{\rho})-U_{23}(\{\bar{\rho})]$.
Using the data {\bf Table 1.}, it is possible to calculate the energy surface of an elementary process, i.e. construct the metric function $\overline{g}(\{\bar{\rho}\})$ (see Fig. 4), which is a necessary condition for 
the complete formulation of a mathematical problem and its preparation for numerical modeling.
\begin{table}[h!]
\centering
\begin{tabular}{||c c c c||} 
 \hline
 $U^{(0)}_{12}=U^{(0)}_{13}=U^{(0)}_{23}$	& $\qquad b_{12}=b_{13}=b_{23}$ & \qquad$ r^0_{12}=r^0_{13}=r^0_{23}$&\qquad   $E$ \\  
 \hline 
 1 &\quad\,\, 0.25 &\quad\,\, 2.0 &\qquad 2.5 \\ 
 1 &\quad\,\, 0.25 &\quad\,\, 2.0 &\qquad 3.5 \\
 \hline
\end{tabular}
\caption{Parameters for determining the total interaction potential of a model three-body system.}
\label{table:1}
\end{table}
 \begin{figure}
\includegraphics[width=45mm]{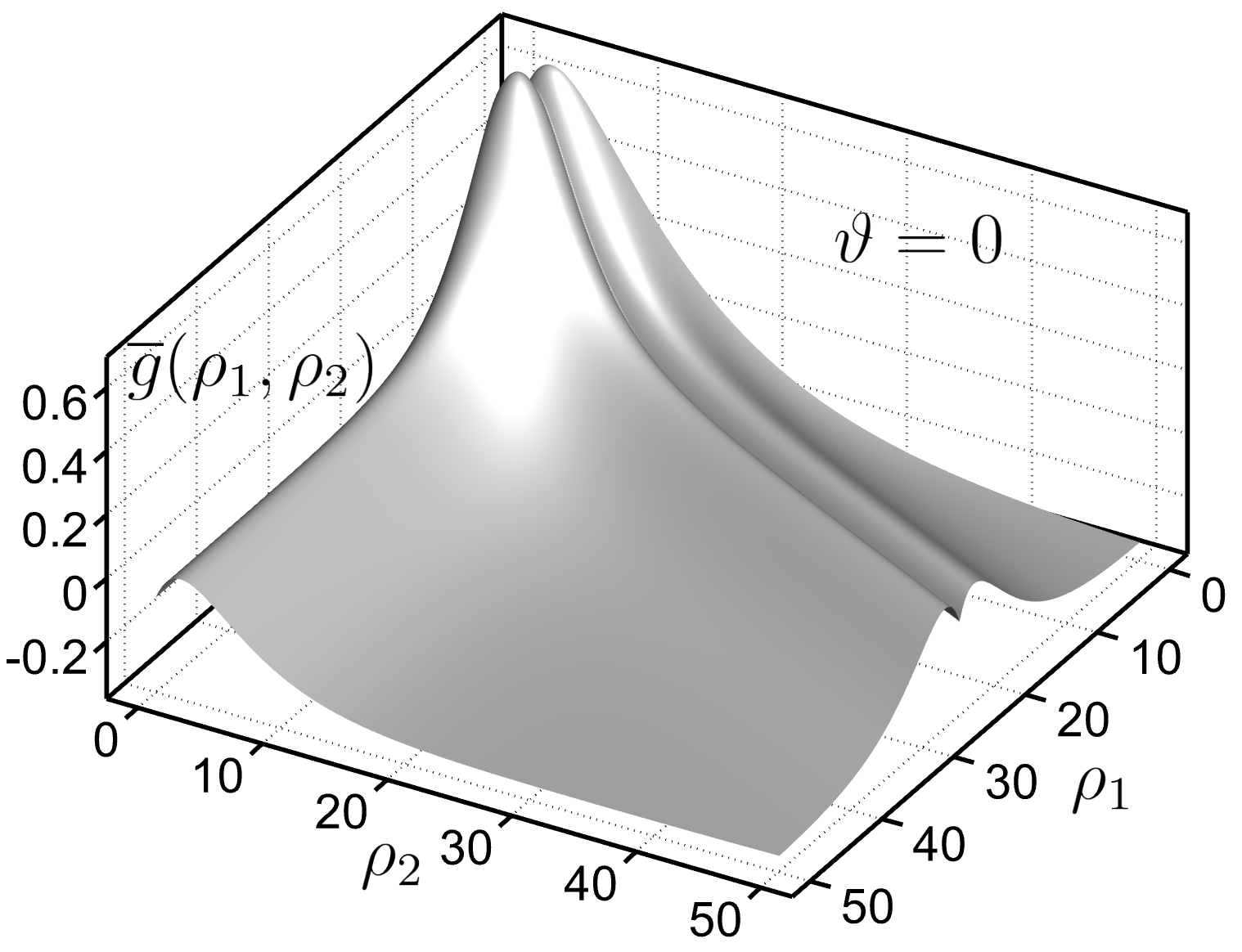}
\includegraphics[width=45mm]{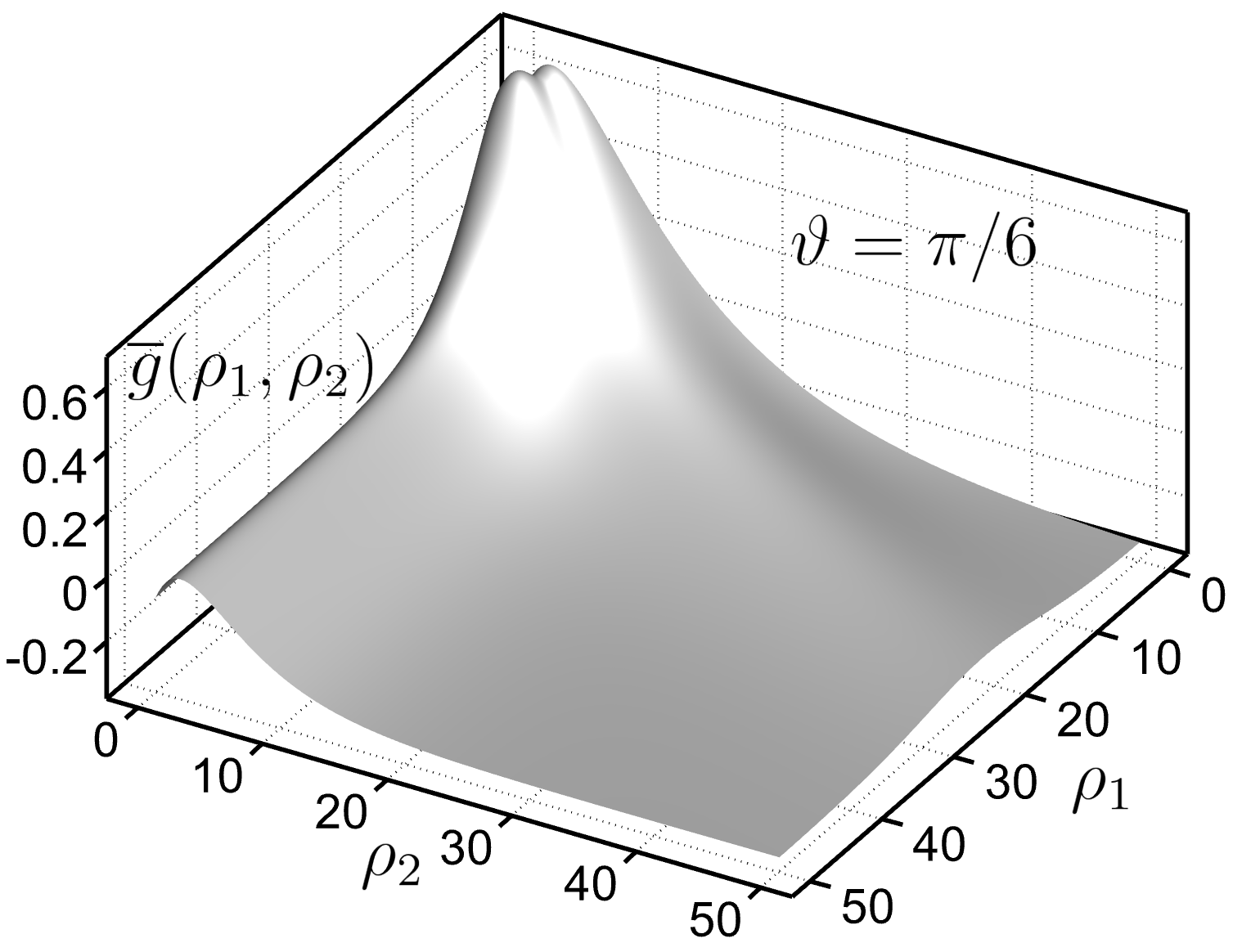}
\includegraphics[width=45mm]{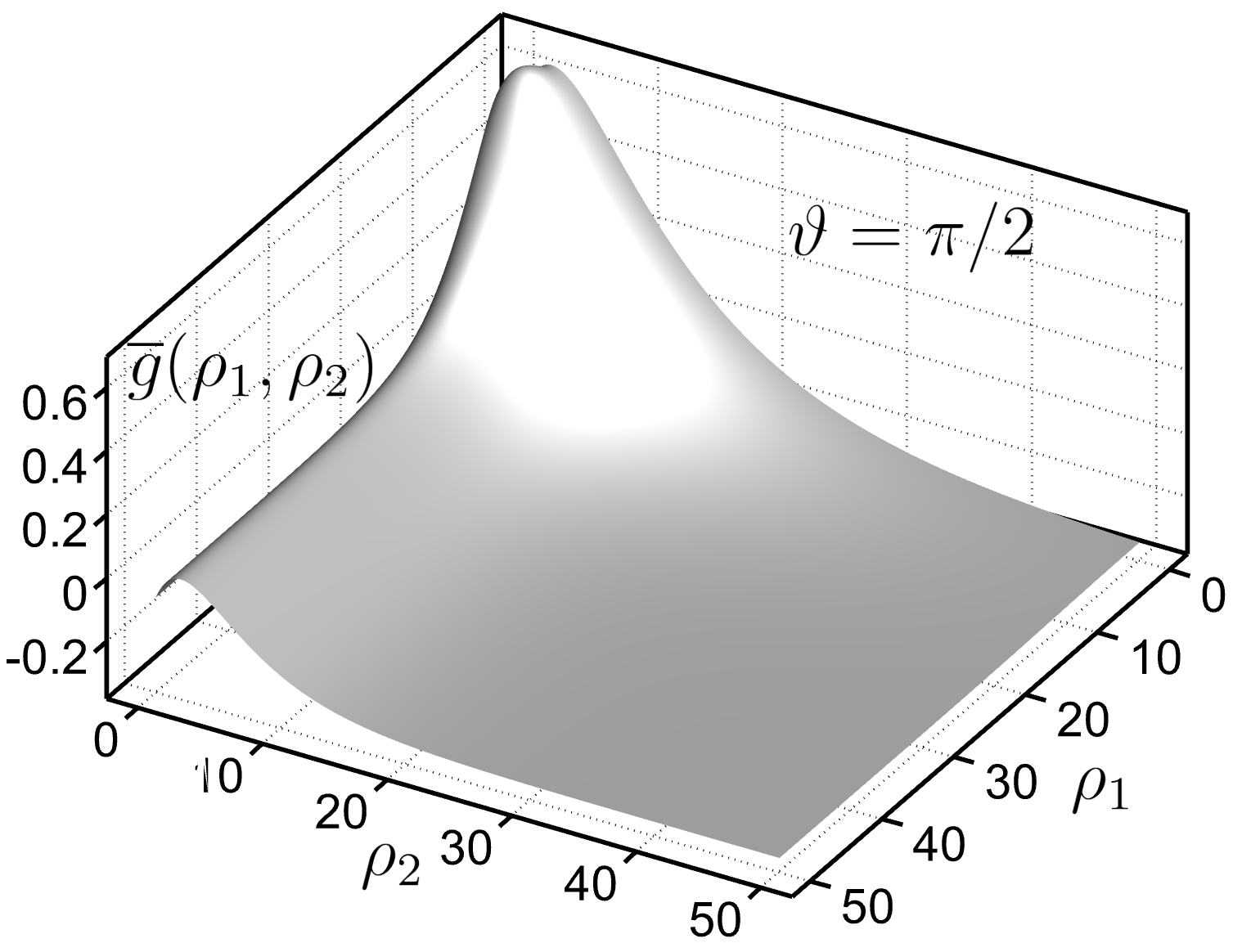}
\caption{\emph{Energy surface of interaction particles for three different scattering angles. Recall that $\rho_3$ in Jacobi coordinates determines the scattering angle, i.e. $\rho_3=\vartheta$ (see Fig. 2).}}
\label{Fig4}
\end{figure}
 Recall that the energy surface plots with data from the second row of {\bf Table 1.} are not given here, since they are similar to the previous one and only are slightly shifted upward due to the large energy value.

\subsection{Underdetermined algebraic Equations and $3D$-manifolds generated by them}
As we have seen, a system of underdetermined algebraic Equations (\ref{q1.04}) plays a key role in transformations between local and global coordinate systems. It is convenient to study this system by writing it 
in a dimensionless form, in which all the geometric and topological properties of the manifolds generated by it are preserved:
\begin{eqnarray}
\bar{\alpha}^2_1+\bar{\beta}_1^2+\bar{\lambda}_1^2=1,\qquad \bar{\alpha}_1\bar{\alpha}_2+\bar{\beta}_1\bar{\beta}_2+\bar{\lambda}_1\bar{\lambda}_2=0,
\nonumber\\
\bar{\alpha}^2_2+\bar{\beta}_2^2+\bar{\lambda}_2^2=1,\qquad \bar{\alpha}_1\bar{\alpha}_3+\bar{\beta}_1\bar{\beta}_3+\bar{\lambda}_1\bar{\lambda}_3=0,
\nonumber\\
\bar{\alpha}^2_3+\bar{\beta}_3^2+\bar{\lambda}_3^2=1,\qquad   \bar{\alpha}_2\bar{\alpha}_3+\bar{\beta}_2\bar{\beta}_3+\bar{\lambda}_2\bar{\lambda}_3=0,
\label{q3.0n1}
\end{eqnarray} 
where $\bar{\alpha}_i=\bar{\alpha}_i/g(\{\bar{x}\})$, $\bar{\beta}_i=\bar{\beta}_i/g(\{\bar{x}\})$ and $\bar{\lambda}_i=\bar{\lambda}_i/g(\{\bar{x}\})$, where $i=1,2,3.$

First, the values for the selected triple of parameters are set. By arbitrarily setting their values from the interval $(-1,+1)$, we can determine the values of the remaining six parameters. Solutions to each triple of 
parameters generate three-dimensional manifolds, the number of which, as indicated above, is equal to 84. Note that we do not consider possible complex solutions to system (\ref{q3.0n1}). Depending on the initial 
triples of parameters specified, the methods for finding the values of the remaining parameters differ significantly from each other. Let us briefly consider the case of specifying $(\bar{\alpha}_1, \bar{\alpha}_2, \bar{\alpha}_3)$, 
which requires an iterative approach. Note that the calculation algorithm itself belongs to the family of gradient methods - the method of conjugate directions.

After a random selection of parameter values $(\bar{\alpha}_1,\bar{\alpha}_2,\bar{\alpha}_3)$, the system of Equations (\ref{q3.0n1}) is completely determined,, i.e. becomes an algebraic system of six 
Equations with six unknowns  $(\bar{\beta}_1,\bar{\beta}_2,\bar{\beta}_3;\bar{\lambda}_1,\bar{\lambda}_2,\bar{\lambda}_3)$ that can be solved exactly. In particular, from the Equations (\ref{q3.0n1}) one can obtain:
\begin{eqnarray}
g_1=\,\bar{\beta}_1^2+\bar{\lambda}_1^2-(1-\bar{\alpha}^2_1)\,=\bar{\beta}_1^2+\bar{\lambda}_1^2-C_1,\qquad\,\,\, C_1=1-\bar{\alpha}^2_1,
\nonumber\\
g_2=\,\bar{\beta}_2^2+\bar{\lambda}_2^2-(1-\bar{\alpha}^2_2)\,=\bar{\beta}_2^2+\bar{\lambda}_2^2-C_2,\qquad\,\,\, C_2=1-\bar{\alpha}^2_2,
\nonumber\\
g_3=\,\bar{\beta}_3^2+\bar{\lambda}_3^2-(1-\bar{\alpha}^2_3)\,=\bar{\beta}_3^2+\bar{\lambda}_3^2-C_3,\qquad\,\,\,  C_3=1-\bar{\alpha}^2_3,
\nonumber\\
g_4= \bar{\alpha}_1\bar{\alpha}_2+\bar{\beta}_1\bar{\beta}_2+\bar{\lambda}_1\bar{\lambda}_2=\bar{\beta}_1\bar{\beta}_2+\bar{\lambda}_1\bar{\lambda}_2+C_4, \quad C_4=\bar{\alpha}_1\bar{\alpha}_2,
\nonumber\\
g_5=\bar{\alpha}_1\bar{\alpha}_3+\bar{\beta}_1\bar{\beta}_3+\bar{\lambda}_1\bar{\lambda}_3=\bar{\beta}_1\bar{\beta}_3+\bar{\lambda}_1\bar{\lambda}_3+C_5,   \quad C_5= \bar{\alpha}_1\bar{\alpha}_3,
\nonumber\\
g_6= \bar{\alpha}_2\bar{\alpha}_3+\bar{\beta}_2\bar{\beta}_3+\bar{\lambda}_2\bar{\lambda}_3=\bar{\beta}_2\bar{\beta}_3+\bar{\lambda}_2\bar{\lambda}_3+C_6, \quad C_6= \bar{\alpha}_2\bar{\alpha}_3.
\label{q3.0nz1}
\end{eqnarray} 
For further research, we introduce new notations:
\begin{equation*}
\mathbf{g} = \left(
\begin{array}{ccc}
g_1  \\
g_2 \\
\vdots \\                                       
g_6
\end{array} \right),
\qquad
\mathbf{x} = \left(
\begin{array}{ccc}
\bar{\beta}_1  \\
\bar{\beta}_2 \\
\vdots \\
\bar{\lambda}_3
\end{array} \right). 
\end{equation*}
We set the objective function $\mathit{K}(\mathbf{x})$:
\begin{equation}
\mathit{K}(\mathbf{x})=\sum_{i=1}^6[g_i(\mathbf{x})]^2=\bigl(\mathbf{g}\cdot\mathbf{g}\bigr).
\label{q3.0nz4}
\end{equation}
Let us define the gradient of the function $\mathit{K}(\mathbf{x})$:
\begin{equation*}
\nabla \mathit{K}(\mathbf{x})= \left(
\begin{array}{ccc}
\frac{\partial\mathit{K}}{\partial x_1} \\
{}\\
\frac{\partial\mathit{K}}{\partial x_2} \\
\vdots \\                                       
\frac{\partial\mathit{K}}{\partial x_6}
\end{array} \right)
=2\left(
\begin{array}{ccc}
2g_1\bar{\beta}_1+g_4\bar{\beta}_2+g_5\bar{\beta}_3 \\
2g_2\bar{\beta}_2 +g_4\bar{\beta}_1+g_6\bar{\beta}_3\\
2g_3\bar{\beta}_3+g_5\bar{\beta}_1+g_6\bar{\beta}_2 \\
2g_1\bar{\lambda}_1+g_4\bar{\lambda}_2+g_5\bar{\lambda}_3\\
2g_2\bar{\lambda}_2+g_4\bar{\lambda}_1+g_6\bar{\lambda}_3\\
2g_3\bar{\lambda}_3+g_5\bar{\lambda}_1+g_6\bar{\lambda}_2\\
\end{array} \right)=2W'\mathit{K}(\mathbf{x}), 
\end{equation*}
where $W'$ is the transposed Jacobian matrix $W$:
\begin{equation*}
W= \left(
\begin{array}{cccccc}
2\bar{\beta}_1 & 0 & 0 & 2\bar{\lambda}_1 & 0 & 0 \\
0 & 2\bar{\beta}_2 & 0 & 0 & 2\bar{\lambda}_2 & 0\\
0 & 0 & 2\bar{\beta}_3 & 0 & 0 & 2\bar{\lambda}_3 \\                                       
\bar{\beta}_2 & \bar{\beta}_1 & 0 & \bar{\lambda}_2 & \bar{\lambda}_1 &0\\
\bar{\beta}_3 & 0 & \bar{\beta}_1 & \bar{\lambda}_3 &0 & \bar{\lambda}_1\\
0 & \bar{\beta}_3 & \bar{\beta}_2 & 0 & \bar{\lambda}_3 & \bar{\lambda}_2\\
\end{array} \right),
\qquad
W'= \left(
\begin{array}{cccccc}
2\bar{\beta}_1 & 0 & 0 & \bar{\beta}_2 & \bar{\beta}_3 & 0 \\
0 & 2\bar{\beta}_2 & 0 & \bar{\beta}_1 & 0 & \bar{\beta}_3\\
0 & 0 & 2\bar{\beta}_3 & 0 & \bar{\beta}_1 & \bar{\beta}_2 \\                                       
2\bar{\lambda}_1 & 0 & 0 & \bar{\lambda}_2 & \bar{\lambda}_3 &0\\
0 & 2\bar{\lambda}_2 & 0 & \bar{\lambda}_1 &0 & \bar{\lambda}_3\\
0 & 0 & 2\bar{\lambda}_3& 0 & \bar{\lambda}_1 & \bar{\lambda}_2\\
\end{array} \right).
\end{equation*}
In iterative methods, the main difficulty usually lies in the correct choice of the initial approximation. For the general case, we do not have a clearly defined criterion for setting such initial values. 
Therefore, we set the initial approximation $\mathbf{x}^{(0)}$ as follows:
\begin{equation}
\mathbf{x}^{(0)} = \left(
\begin{array}{ccc}
\bar{\beta}_1^{(0)}  \\
\bar{\beta}_2^{(0)} \\
\vdots \\                                       
\bar{\lambda}_3^{(0)}
\end{array} \right),
\qquad
\begin{array}{ccc}
\bar{\beta}_1^{(0)} = \pm\sqrt{(1-\bar{\alpha}_1^2)/2}, \qquad \bar{\lambda}_1^{(0)}=\pm\bar{\beta}_1^{(0)},
\\
\bar{\beta}_2^{(0)} = \pm\sqrt{(1-\bar{\alpha}_2^2)/2}, \qquad \bar{\lambda}_2^{(0)}=\pm\bar{\beta}_2^{(0)},
\\
\bar{\beta}_3^{(0)} = \pm\sqrt{(1-\bar{\alpha}_3^2)/2}, \qquad\, \bar{\lambda}_3^{(0)}=\pm\bar{\beta}_3^{(0)}.
\end{array}
\label{qn3.0n1}
\end{equation}
The signs in the initial approximation $\mathbf{x}^{(0)}$ of system (\ref{qn3.0n1}) are set randomly, but so that for the first three approximations and the next three values the signs are not 
the same. This is due to the form of the last three equations of system (\ref{q3.0nz1}). 

Finally, we can set the calculation algorithm:
\begin{equation}
\mathbf{x}^{(n+1)}=\mathbf{x}^{(n)}+\delta_n\mathbf{p}^{(n)},\quad \mathbf{p}^{(n)}=-\nabla \mathit{K}(\mathbf{x}^{(n)})+\gamma_{n-1}\mathbf{p}^{(n-1)},\qquad n=1,2,...
\label{q6.t01}
\end{equation}
where $\mathbf{p}^{(n)}$  is the direction of descent,  $\delta_{n}$ is the step of descent and the coefficient $\gamma_{n}$, which is defined below in step 4.

1. We set the initial approximation $\mathbf{p}^{(0)}$:
$$
\mathbf{p}^{(0)}=-\nabla \mathit{K}(\mathbf{x}^{(0)}),\qquad n=0,
$$

2. Determine the next value $\mathbf{x}^{(1)}$:
$$
\mathbf{x}^{(1)}=\mathbf{x}^{(0)}+\delta_0\mathbf{p}^{(0)}.
$$
The value of the descent step $\delta_n\,( n \geqslant 0)$ is searched numerically until the condition is met:
$$
\mathit{K}(\mathbf{x}^{(n)}+\delta_n\mathbf{p}^{(n)})<\mathit{K}(\mathbf{x}^{(n)}),\qquad \nabla\mathit{K}(\mathbf{x}^{(n)})=2W'(\mathbf{x}^{(n)})\mathbf{g}(\mathbf{x}^{(n)}). 
$$

3. We calculate the term $\mathbf{p}^{(1)}$:
$$
\mathbf{p}^{(1)}=-\nabla \mathit{K}(\mathbf{x}^{(1)})+\gamma_0\mathbf{p}^{(0)}.
$$

4. The coefficients $\gamma_{n-1}$ are calculated using the formula:
$$
\gamma_{n-1}=\frac{\bigl[\bigl(\nabla\mathit{K}(\mathbf{x}^{(n)}),\nabla\mathit{K}(\mathbf{x}^{(n)})\bigr)-\nabla\mathit{K}(\mathbf{x}^{(n-1)})\bigr]}{|\nabla\mathit{K}(\mathbf{x}^{(n-1)})|^2},\qquad n\geqslant1,
$$

5. For subsequent $n$ we repeat steps 2-4 until the condition $|\nabla\mathit{K}(\mathbf{x}^{(n)})|\leqslant\varepsilon_1$, where $\varepsilon_1$ is the specified accuracy.

 After obtaining a chain of approximations $\mathbf{x}^{(n)}$  upon reaching the end condition of iterations 5, we refine the initial initial data $(\bar{\alpha}_1,\bar{\alpha}_2,\bar{\alpha}_3)$.
Note that the following notations are used above:
$$
|\mathbf{y}|^2=\sum^6_{i=1}y_i^2,\qquad (\mathbf{a},\mathbf{b})=\sum a_i b_i.
$$
The refined initial approximation is obtained from the correction of the initial data parameters $(\bar{\alpha}_1,\bar{\alpha}_2,\bar{\alpha}_3)$ according to the original scheme (\ref{qn3.0n1}). This is justified by 
the fact that initially the value of the triple of parameters is chosen randomly.

Then we carry out the next stage of iterations using the described method 1-5. The end of such an outer iteration occurs when the condition is met;
$
|\nabla\mathit{K}(\mathbf{x}^{(n)})|\leqslant\varepsilon_2.
$

In order to avoid excessive iteration steps or looping, an additional limitation on the number of iterations is introduced. Recall that during the calculations the values were specified $\varepsilon_1=10^{-9}$
and $\varepsilon_2=10^{-18}$. Such values make it possible to determine the roots of the algebraic system (\ref{q3.0n1}) with an accuracy at which  $g_i \approx 10^{-10}$, $i=4,5,6$ and $g_i \approx 10^{-16}$,
$ i=1,2,3$ for the system (\ref{q3.0nz1}).

The solution of Equations system (\ref{q3.0n1}) for the original triplets $(\bar{\beta}_1,\bar{\beta}_2,\bar{\beta}_3)$ and $(\bar{\lambda}_1,\bar{\lambda}_2,\bar{\lambda}_3)$ is similar to the method described above. To illustrate the calculations, we present here two complete members of the families of manifolds $\mathcal{A}$ and $\mathcal{B}$ (see Fig. 4 and Fig. 5), on which we will further study the dynamics of the classical  three-body system.
\begin{figure}
\includegraphics[width=45mm]{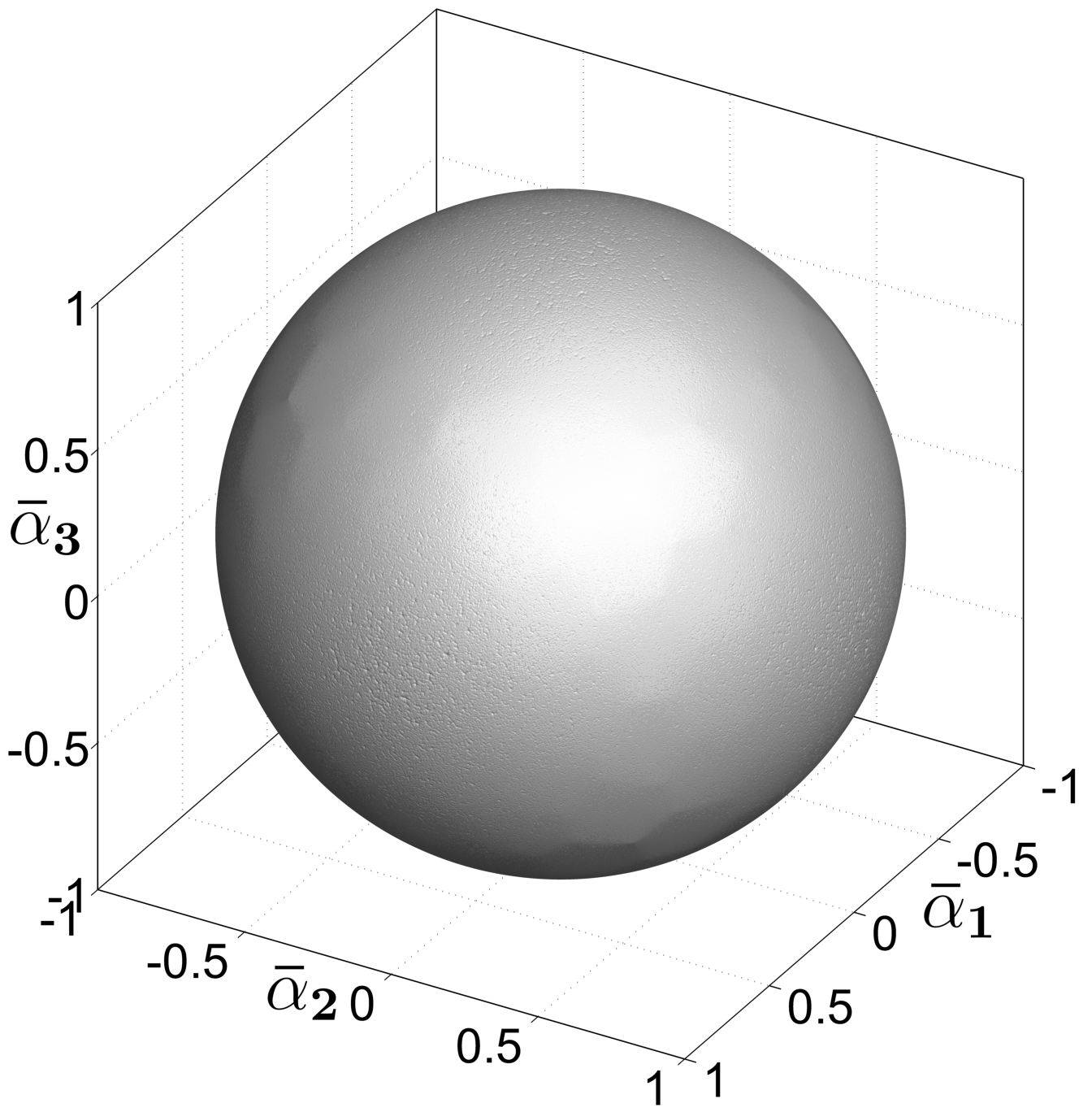}
\includegraphics[width=45mm]{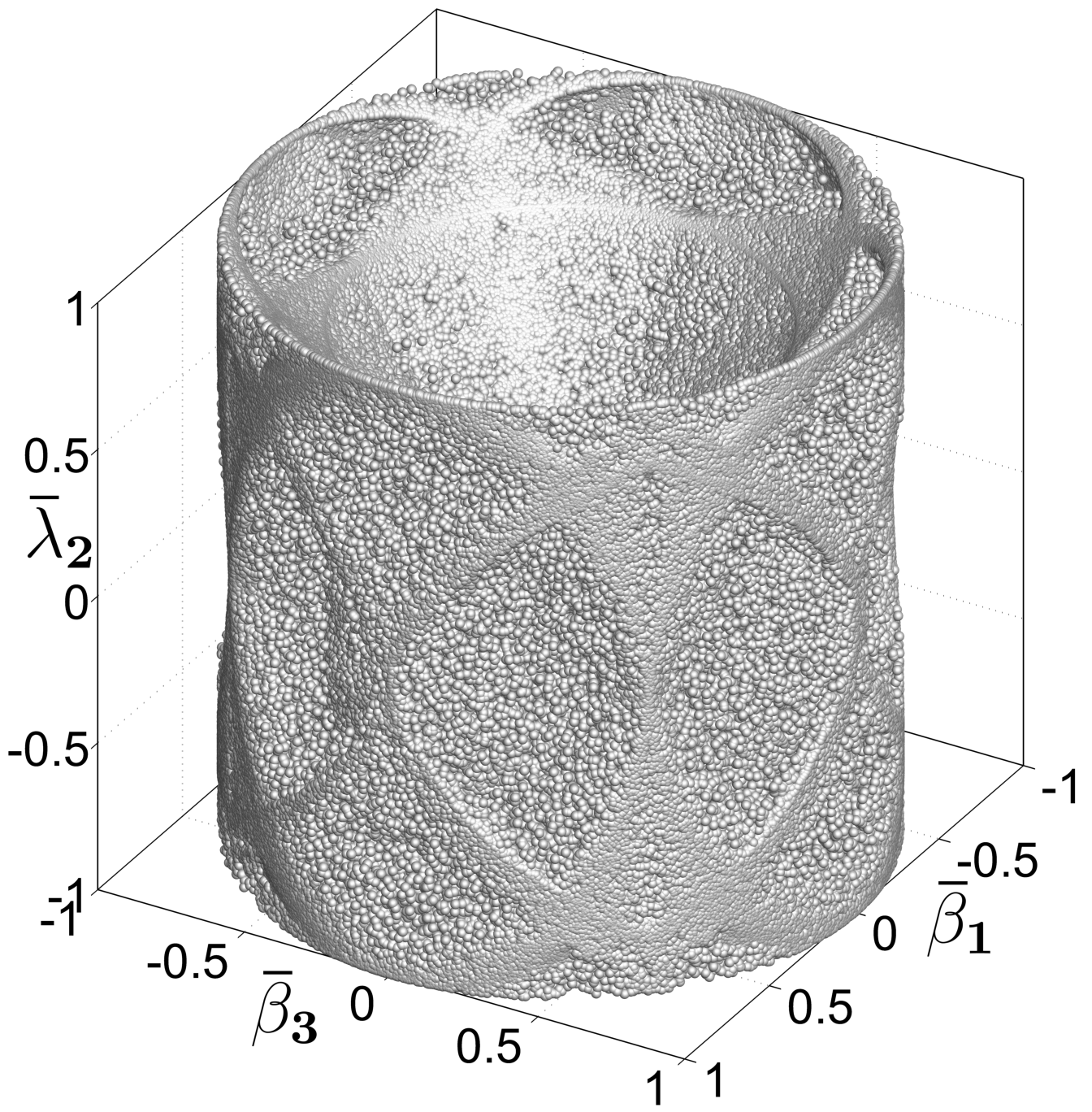}
\includegraphics[width=45mm]{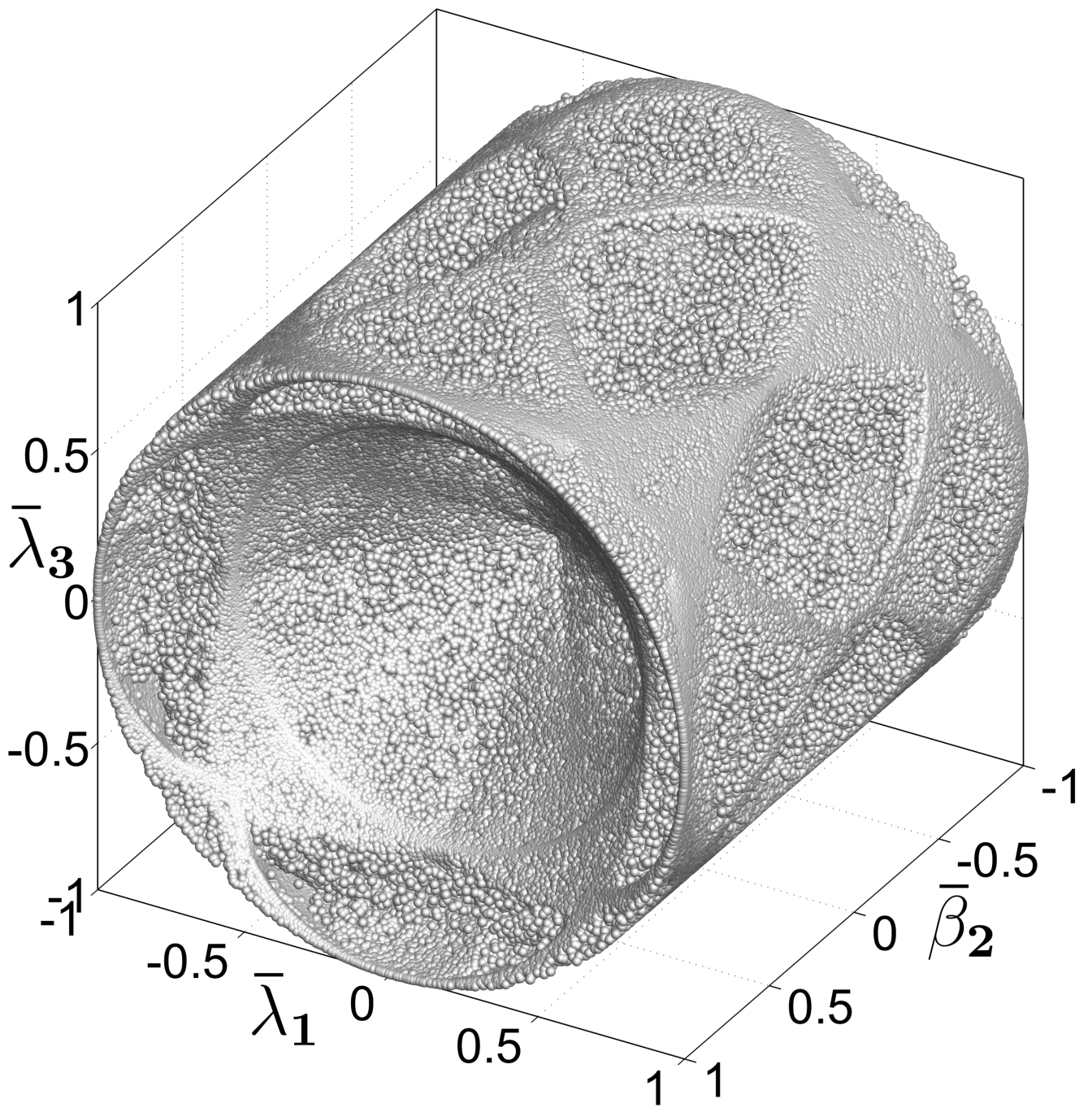}
\caption{A manifold of the family $\mathcal{A}$, which has the form $\mathcal{R}^{(3)}_{(\beta_1,\lambda_2,\beta_3)}$ (sphere) and two additional manifolds surrounding it from  left to right 
$\mathcal{R}^{( 3) }_{(\beta_1,\lambda_2,\beta_3)}$ and $ \mathcal{R}^{(3)}_{(\lambda_1,\beta_2,\lambda_3)}$. Combining these manifolds by a direct product, we obtain a complete member of the family $\mathcal{A}$, which can be represented in the following form
$\mathcal{R}^{(9)}_{\mathcal{A}_1}=\mathcal{R}^{(3)}_{(\beta_1,\lambda_2,\beta_3)}\times\mathcal{R}^{( 3) }_{(\beta_1,\lambda_2,\beta_3)}\times\mathcal{R}^{(3)}_{(\lambda_1,\beta_2,\lambda_3)}$.}
\label{Fig5}
\end{figure}
\begin{figure}
\includegraphics[width=45mm]{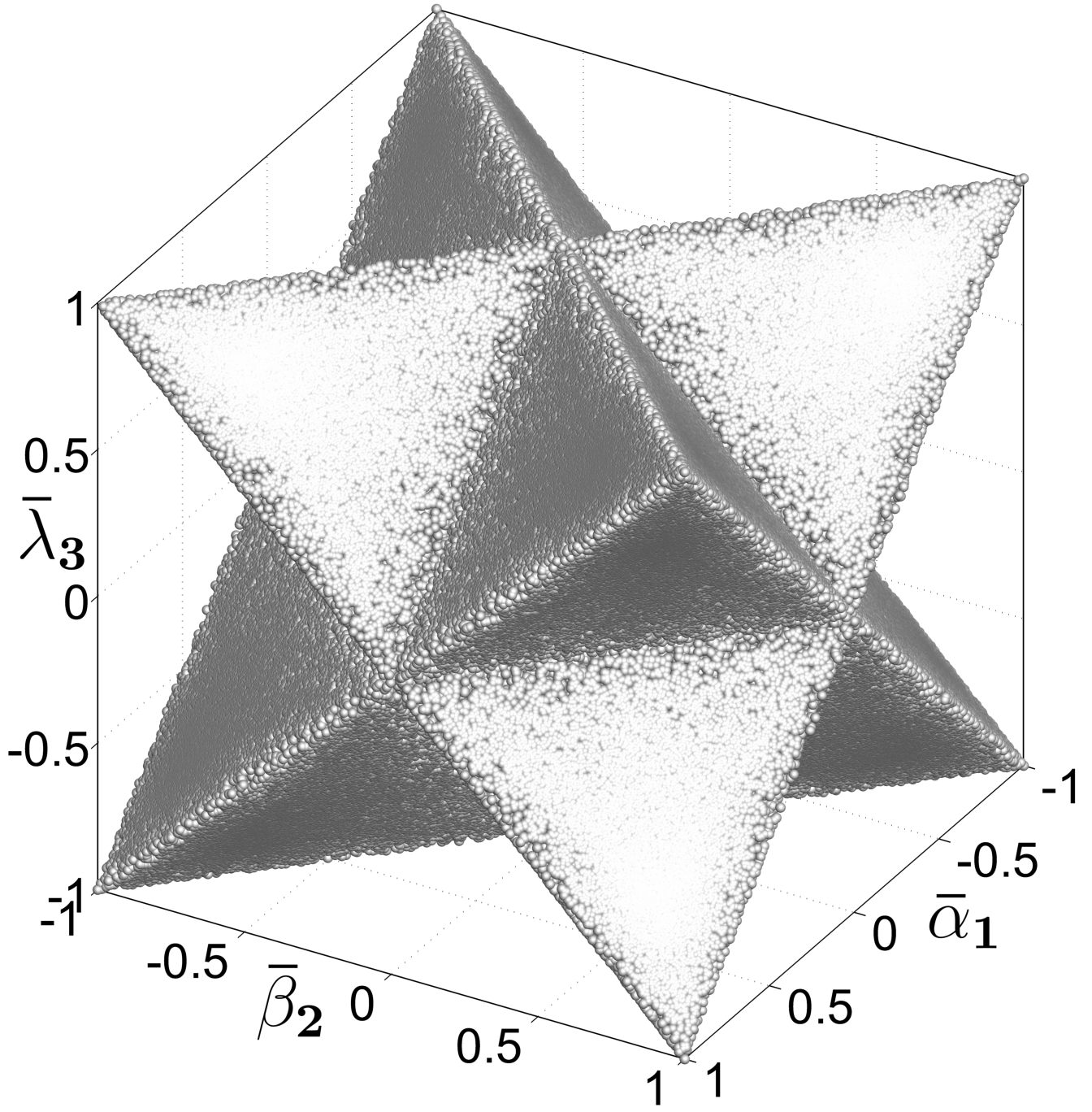}
\includegraphics[width=45mm]{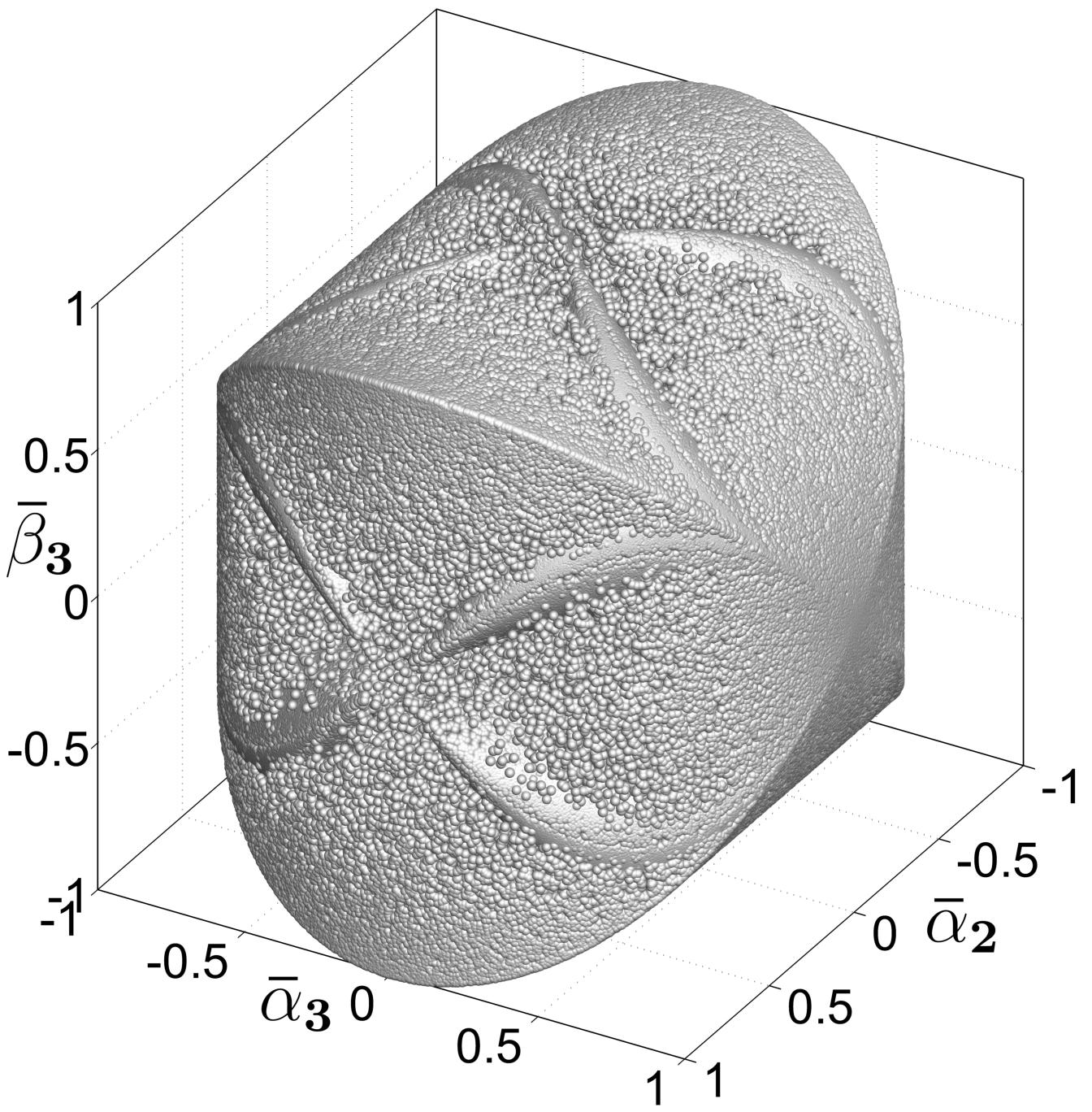}
\includegraphics[width=45mm]{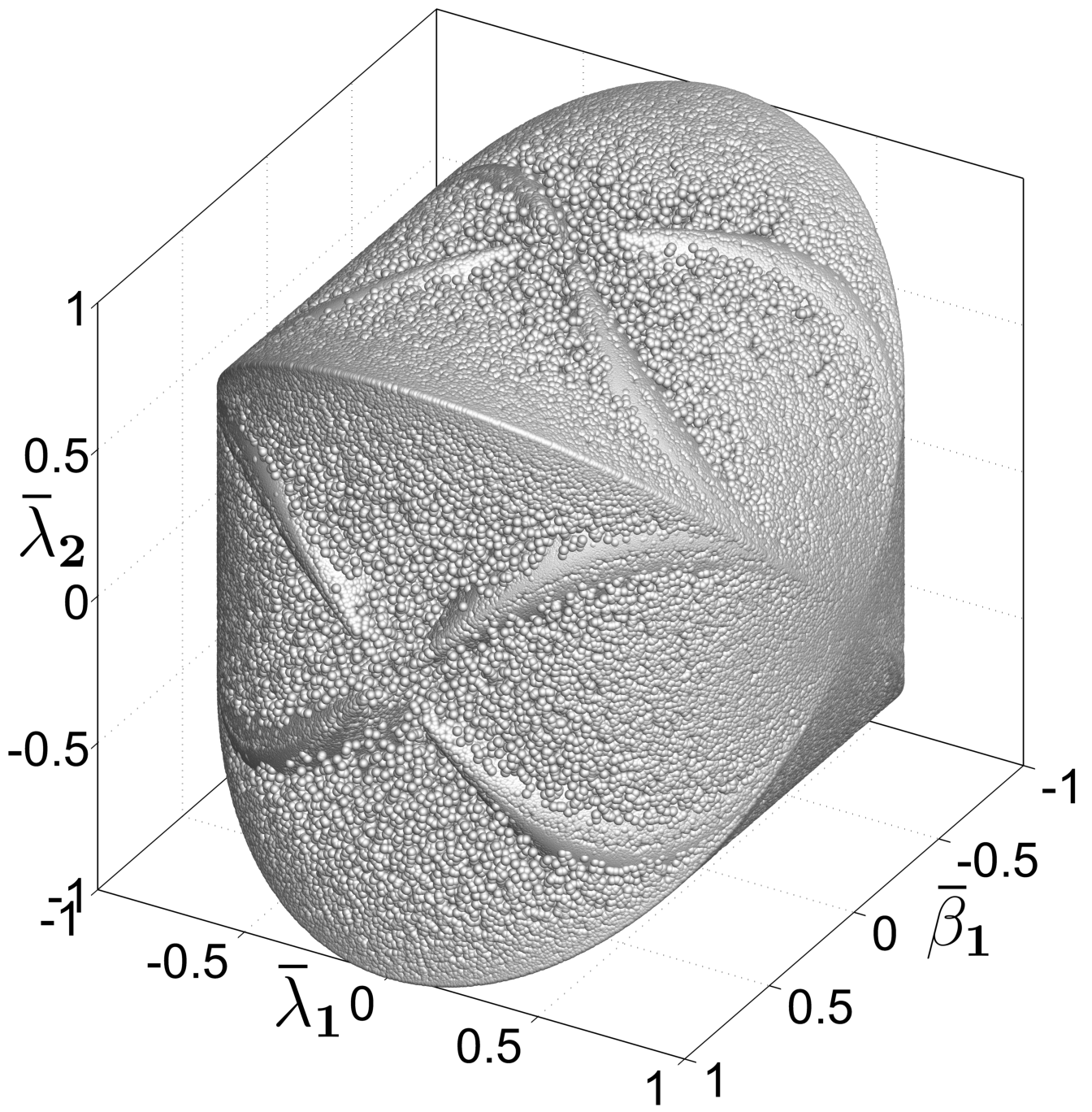}
\caption{A manifold of the family $\mathcal{B}$, which has the form  $\mathcal{R}^{(3)}_{(\alpha_1,\beta_2,\lambda_3)}$ (two three-dimensional pyramids fastened together) and two additional manifolds surrounding  it from  left to right  $\mathcal{R}^{( 3) }_{(\alpha_2,\alpha_3,\beta_3)}$ and  $ \mathcal{R}^{(3)}_{(\beta_1,\lambda_1,\lambda_2)}$. Combining these manifolds by a direct product, we obtain a complete member of  the family $\mathcal{B}$, which can be represented in the following form $ \mathcal{R}^{(9)}_{\mathcal{B}_1}=\mathcal{R}^{(3)}_{(\alpha_1,\beta_2,\lambda_3)}\times \mathcal{R}^{(3)}_{(\alpha_2,\beta_3,\lambda_1)}\times
 \mathcal{R}^{(3)}_{(\alpha_3,\beta_1,\lambda_2)}$.}
\label{Fig6}
\end{figure}

\subsection{Mathematical algorithm for modeling a trajectory problem}

Let us rewrite the system of differential equations (\ref{q1.01}) in a local coordinate system. In general, it can be represented as follows:
\begin{equation}
\frac{d\bf{Y}}{dt}={\bf\Phi}(t;u,v,w,x^1,x^2,x^3),
\label{qcV.01}
\end{equation}
where $u=\dot{x}^1$,\,\,$v=\dot{x}^2$ and $w=\dot{x}^3$, in addition,  the following notations are introduced:
\begin{equation}
\bf{Y} = \left(
\begin{array}{ccc}
u \\
v \\
w \\                                       
x^1\\
x^2\\
x^3\\
\end{array} \right),
\qquad
{\bf\Phi}= \left(
\begin{array}{cccccc}
\Phi_1\\
\Phi_2\\
\Phi_3\\
\Phi_4\\
\Phi_5\\
\Phi_6\\
\end{array} \right)
= \left(
\begin{array}{cccccc}
a_1(u^2-v^2-w^2-\Lambda^2) +2u(a_2v+a_3w)\\ 
a_2(v^2-w^2-u^2-\Lambda^2)+2v(a_3w+a_1u)\\
a_3(w^2-u^2-v^2-\Lambda^2) +2w(a_1u+a_2v)\\                                       
u\\
v\\
w\\
\end{array} \right).
\label{qcV.02}
\end{equation}
For the system of differential equations (\ref{qcV.01}), the Cauchy problem with initial conditions is posed:
\begin{equation}
{\bf Y}(t_0)={\bf Y}_0.
\label{qV.03}
\end{equation}

According to the Runge–Kutta method, the approximate value of $\bf{Y}_{i+1}$ is calculated using the formulas with a step $\Delta t$:
\begin{equation}
{\bf Y}_{i+1}={\bf Y}_{i}+\Delta{\bf Y}_i, \qquad i=0,1,2,...
\label{qV.04}
\end{equation}
where 
\begin{equation}
\Delta{\bf Y}_i =\frac{1}{6}\bigl[{\bf k}_1^{(j)}+2{\bf k}_2^{(j)}+2{\bf k}_3^{(j)}+{\bf k}_4^{(j)}\bigr], \qquad j=\overline{1,6}.
\label{qV.05}
\end{equation}
 In addition, the following notations are made in (\ref{qV.04}):
\begin{eqnarray}
{\bf k}_1^{(j)}={\Delta t}\cdot{\bf\Phi}^j(t_i, u_i,v_i,w_i,x^1_i,x^2_i,x^3_i),\qquad\qquad\qquad\qquad\qquad\qquad\qquad\qquad\qquad\qquad\qquad
\nonumber\\
{\bf k}_2^{(j)}={\Delta t}\cdot{\bf\Phi}^j \Bigl[t_i+\frac{\Delta t}{2}, u_i+\frac{{\bf k}_1^{(1)}}{2},v_i+\frac{{\bf k}_1^{(2)}}{2},w_i+\frac{{\bf k}_1^{(3)}}{2},x^1_i+\frac{{\bf k}_1^{(4)}}{2}, x^2_i+\frac{{\bf k}_1^{(5)}}{2},x^3_i+\frac{{\bf k}_1^{(6)}}{2}\Bigr],\,\,
\nonumber\\
{\bf k}_3^{(j)}={\Delta t}\cdot{\bf\Phi}^j\Bigl[t_i+\frac{\Delta t}{2}, u_i+\frac{{\bf k}_2^{(1)}}{2},v_i+\frac{{\bf k}_2^{(2)}}{2},w_i+\frac{{\bf k}_2^{(3)}}{2}, x^1_i+\frac{{\bf k}_2^{(4)}}{2},x^2_i+\frac{{\bf k}_2^{(5)}}{2},x^3_i+\frac{{\bf k}_2^{(6)}}{2}\Bigr],\,\,
\nonumber\\
{\bf k}_4^{(j)}={\Delta t}\cdot{\bf\Phi}^j\Bigl[t_i\,+{\Delta t},\, u_i+{{\bf k}_3^{(1)}},\,v_i+\,{\bf k}_3^{(2)},\,w_i+{\bf k}_3^{(3)},\,x^1_i+\,{\bf k}_3^{(4)},\,x^2_i+{\bf k}_3^{(5)}, \,x^3_i+{\bf k}_3^{(6)}\Bigr].\,\,\,
\label{qV.06}
\end{eqnarray}
Note that during calculations the calculation step remains constant, equal to $\Delta t=10^{-4}.$

Solving the system of equations (\ref{qcV.01})-(\ref{qcV.02}) we obtain the trajectory of the center of mass of the three-body system in the local coordinate system. For the final presentation of the results, 
they must be presented in the system of Jacobi variables (global variables). Using differential representations for global coordinates (\ref{q1.03}), we can write the following expressions:
\begin{eqnarray}
\rho_1=\rho_{01}+\,\int_{x^1_0}^{x^1}\alpha_1dx\,+\int_{x^2_0}^{x^2}\alpha_2dy+\int_{x^3_0}^{x^3}\alpha_3dz=\,\rho_{01}+\alpha_1\Delta x^1\,+\alpha_2\Delta x^2+\alpha_3\Delta x^3,
\nonumber\\
\rho_2=\rho_{02}+\int_{x^1_0}^{x^1}\beta_1dx+\int_{x^2_0}^{x^2}\beta_2dy+\int_{x^3_0}^{x^3}\beta_3dz=\rho_{02}+\beta_1\Delta x^1\,+\beta_2\Delta x^2+\,\beta_3\Delta x^3,
\nonumber\\
\rho_3=\rho_{03}+\int_{x^1_0}^{x^1}\lambda_1dx+\int_{x^2_0}^{x^2}\lambda_2dy+\int_{x^3_0}^{x^3}\lambda_3dz=\rho_{03}+\lambda_1\Delta x^1\,+\lambda_2\Delta x^2\,+\lambda_3\Delta x^3,
\label{qV.07}
\end{eqnarray}
where $\Delta x^1=x^1-x_0^1,$\,\,$\Delta x^2=x^2-x_0^2$ and $\Delta x^3=x^3-x_0^3$, in addition, $\{\bar{x}_0\}=(x^1_{0},x^2_{0},x^3_{0})$ and $\{\bar{\rho}_0\}=(\rho_{01},\rho_{02},\rho_{03})$ 
denote the initial coordinate values in the local and global coordinate systems, respectively. 

In the same way, we can define equations for velocities in global coordinate system using increments of velocities in local coordinates:
\begin{equation}
\begin{cases}
\dot{\rho}_1=\dot{\rho}_{01} +\alpha_1\Delta u+\,\alpha_2\Delta v+\alpha_3\Delta w,
\nonumber\\
\dot{\rho}_2=\dot{\rho}_{02}+\beta_1\Delta u+\beta_2\Delta v+\beta_3\Delta w,
\nonumber\\
\dot{\rho}_3=\dot{\rho}_{03}+\lambda_1\Delta u+\lambda_2\Delta v+\lambda_3\Delta w,
\label{qV.08}
\end{cases}
\end{equation}
where $\dot{\rho}_{01},$\, $\dot{\rho}_{02}$ and $\dot{\rho}_{03}$ denote the components of the initial velocity of the center of mass of a three-body system in global coordinates.

The initial value of the coordinate of the phase point in the local system $\{\bar{x}_0\}=(x^1_0,x^2_0,x^3_3)$ at  moment $s = 0$ must correspond to the initial value $\{\bar{\rho}_0\}
=(\rho_{01},\rho_{02},\rho_{03})$ at moment $t = 0$.  Due to the specifics of calculating such evolutionary problems, the initial values $\{\bar{x}\}$ and $\{\dot{\bar{x}}\}$ can be 
taken as the values of the coordinates and velocity of the point of the previous step of the system calculation at the moment $t=t_0+\Delta t$, where $\Delta t$ is the time step for 
the local coordinate system, $t_0$ is the origin of the count at this step.

For definiteness, we will assume that the three bodies have the same masses and at the moment $t = 0$ form an equilateral triangle in the Cartesian coordinate system. In this case, 
the initial values of the Jacobi coordinates will be as follows:
\begin{equation}
{\rho}_1|_{t=0}=\,\rho_{01}\,=\,\frac{\sqrt{3}}{2},\qquad {\rho}_2|_{t=0}\,=\,\rho_{02}\,=\,1,\qquad {\rho}_3 |_{t=0}\,=\,\rho_{03}\,=\,\frac{\pi}{2}.
\label{qV.09}
\end{equation}
\begin{table}[h!]
\centering
\begin{tabular}{||c c c c||} 
 \hline
 $\dot{\rho}_{1} |_{t=0}=\dot{\rho}_{01}\qquad$	& $\dot{\rho}_{2} |_{t=0}=\dot{\rho}_{02}\qquad$ & $ \dot{\rho}_{3} |_{t=0}=\dot{\rho}_{03}\qquad$ & \qquad  $J$ \\  
 \hline 
0.01		&0.01			& 0.10&\qquad 0.30\\ 
0.30		&0.50			& 0.40 &\qquad 0.60\\
1.00		&0.80			&0.60 &\qquad 0.80\\
 \hline
\end{tabular}
\caption{For the initial velocities $\{\dot{\bar{\rho}}\}$ and the moment of inertia of the three-body system $J$ we use the following values. }
\label{table:2}
\end{table}
 It is very important now to synchronize the initial coordinate values in the Jacobi system $\{\bar{\rho}\}\in \mathbb{R}^{(3)}$ and in the local coordinate system $\{\bar{x}\}\in \mathcal{R}^{(3)}_{\mathcal{F}}$,
where $\mathcal{F}=(\mathcal{A}_1;\,\mathcal{B}_1)$ describes two families of manifolds. Using equations (\ref{q1.05})-(\ref{q1.06}), we can write the following transformations for the position and velocity of the 
reduced mass $\mu_0$ in the local coordinate system:
\begin{eqnarray}
x^1|_{t=0}=x^1_0=\breve{\alpha}_1\rho_{01}+\breve{\alpha}_2\rho_{02}+\breve{\alpha}_3\rho_{03}, \qquad \dot{x}^1|_{t=0}=\dot{u}_{0}=\breve{\alpha}_1\dot{\rho}_{01}+\breve{\alpha}_2\dot{\rho}_{02}+\breve{\alpha}_3\dot{\rho}_{03},\,  
\nonumber\\ 
x^2|_{t=0}=x^2_0=\breve{\beta}_1\rho_{01}+\breve{\beta}_2\rho_{02}+\breve{\beta}_3\rho_{03}, \qquad \dot{x}^2|_{t=0}=\dot{v}_{0}= \breve{\beta}_1\dot{\rho}_{01}+\breve{\beta}_2\dot{\rho}_{02}+\breve{\beta}_3\dot{\rho}_{03},
\nonumber\\
x^3|_{t=0}=x^3_0=\breve{\lambda}_1\rho_{01}+\breve{\lambda}_2\rho_{02}+\breve{\lambda}_3\rho_{03},\qquad \dot{x}^3|_{t=0}=\dot{w}_{0}= \breve{\lambda}_1\dot{\rho}_{01}+\breve{\lambda}_2\dot{\rho}_{02}+\breve{\lambda}_3\dot{\rho}_{03}.  
\label{qV.10}
\end{eqnarray}

Thus, now we have all the tools to organize the numerical integration of the trajectory problem in the global Jacobi coordinate system.

\section{Calculation of the internal time of the restricted three-body problem and discussions}
 
If we use the data from the first rows of {\bf Table 1} and {\bf Table 2}, and also consider the initial configuration of the three bodies given in (\ref{qV.09}), then, obviously, the system of bodies will be in a bound state 
throughout the entire time of movement since the binding energy will be greater than the kinetic energy at finite distances (restricted three-body problem). In particular, by calculating the internal time of a dynamical 
system on complete members of the families of manifolds $\mathcal{A}_1$ and  $\mathcal{B}_1$ (see Fig. 5 and 6) using the algorithm given in Section 5. , we find their three-dimensional images (see Fig. 7).
\begin{figure}
\includegraphics[width=40mm]{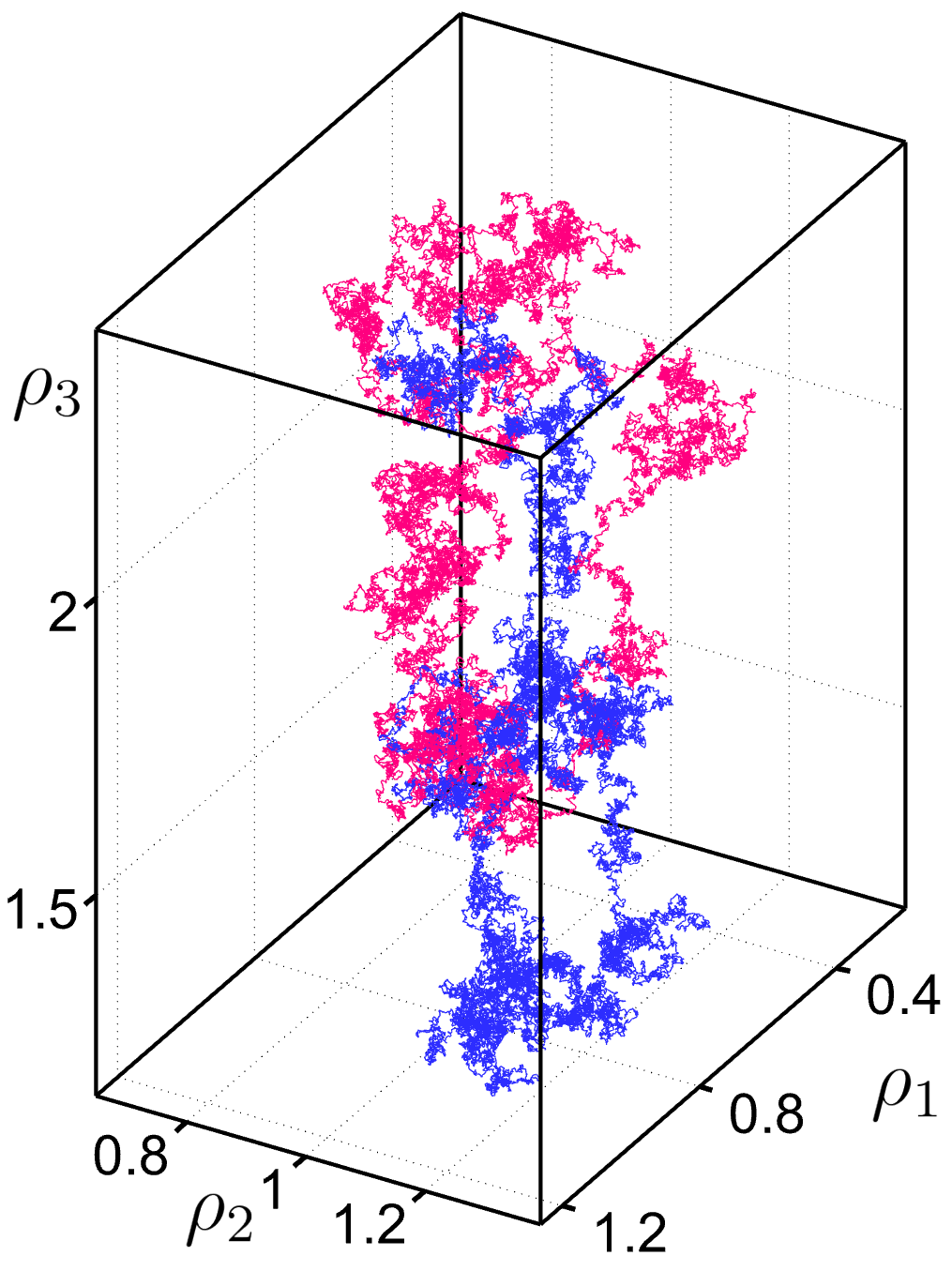}
\includegraphics[width=45mm]{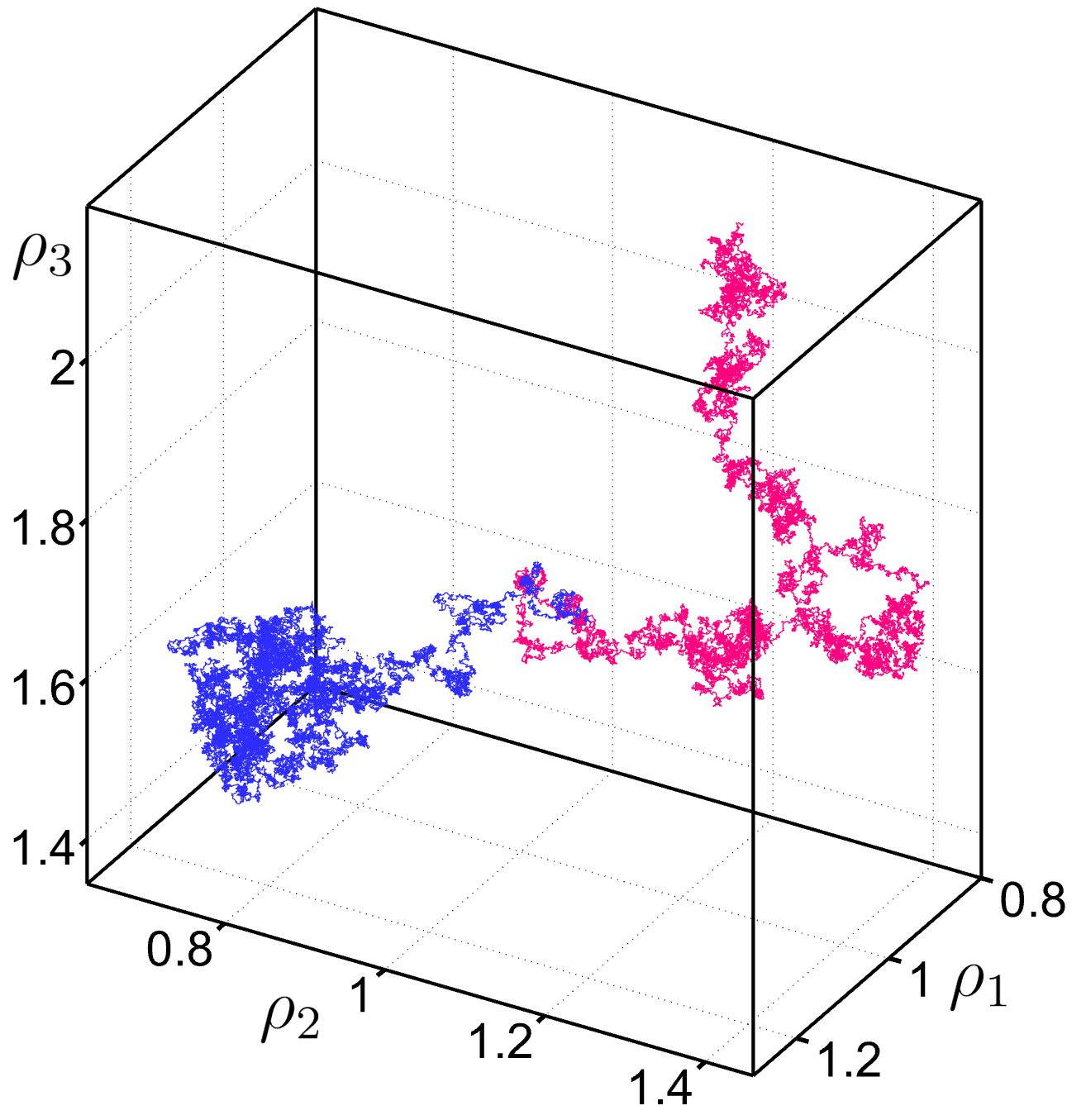}
\includegraphics[width=50mm]{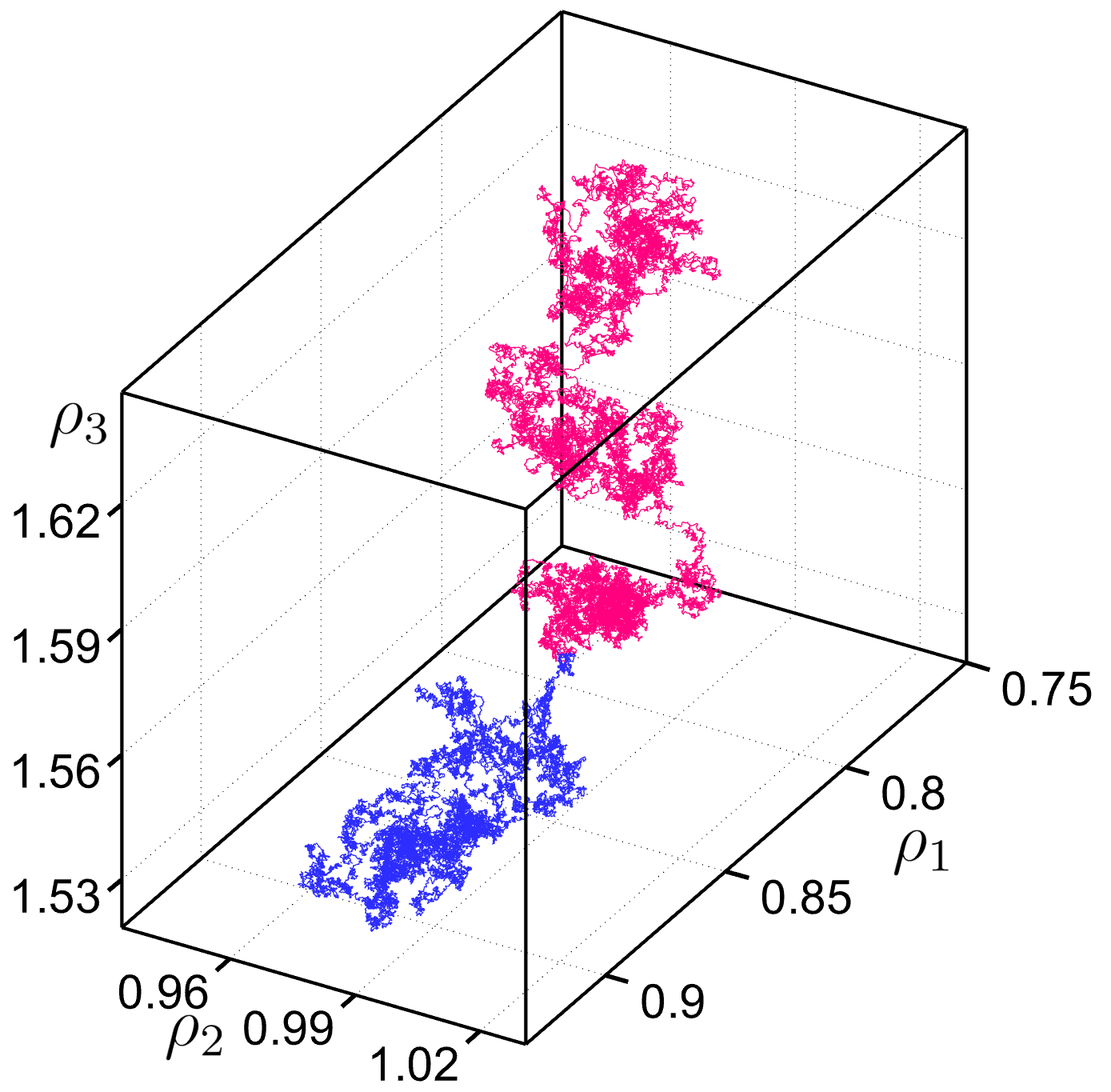}
\caption{\emph{Internal time of three particles for three different initial data on two different complete terms of the manifolds $\mathcal{A}_1$ and $\mathcal{B}_1$. On the plots, blue and red colors indicate internal 
times that were calculated on the manifolds $\mathcal{A}_1$ and $\mathcal{B}_1$, respectively. Each point of internal time, if projected onto the coordinate axes, determines the configuration of three particles at a given moment.}}
\label{Fig7}
\end{figure}

Analysis of the calculations shows that when using the above initial data, the resulting trajectories indeed describe a restricted three-body problem, however, their nature is chaotic. To make sure that the system's motion 
is truly chaotic, Lyapunov exponents are calculated for two trajectories whose initial data are close and differ by $10^{-2}$. Recall that geodesic trajectories are calculated both on the manifold $\mathcal{A}_1$
and on $\mathcal{B}_1$, see Fig.s 8 and 9.
\begin{figure}
\quad
\includegraphics[width=55mm]{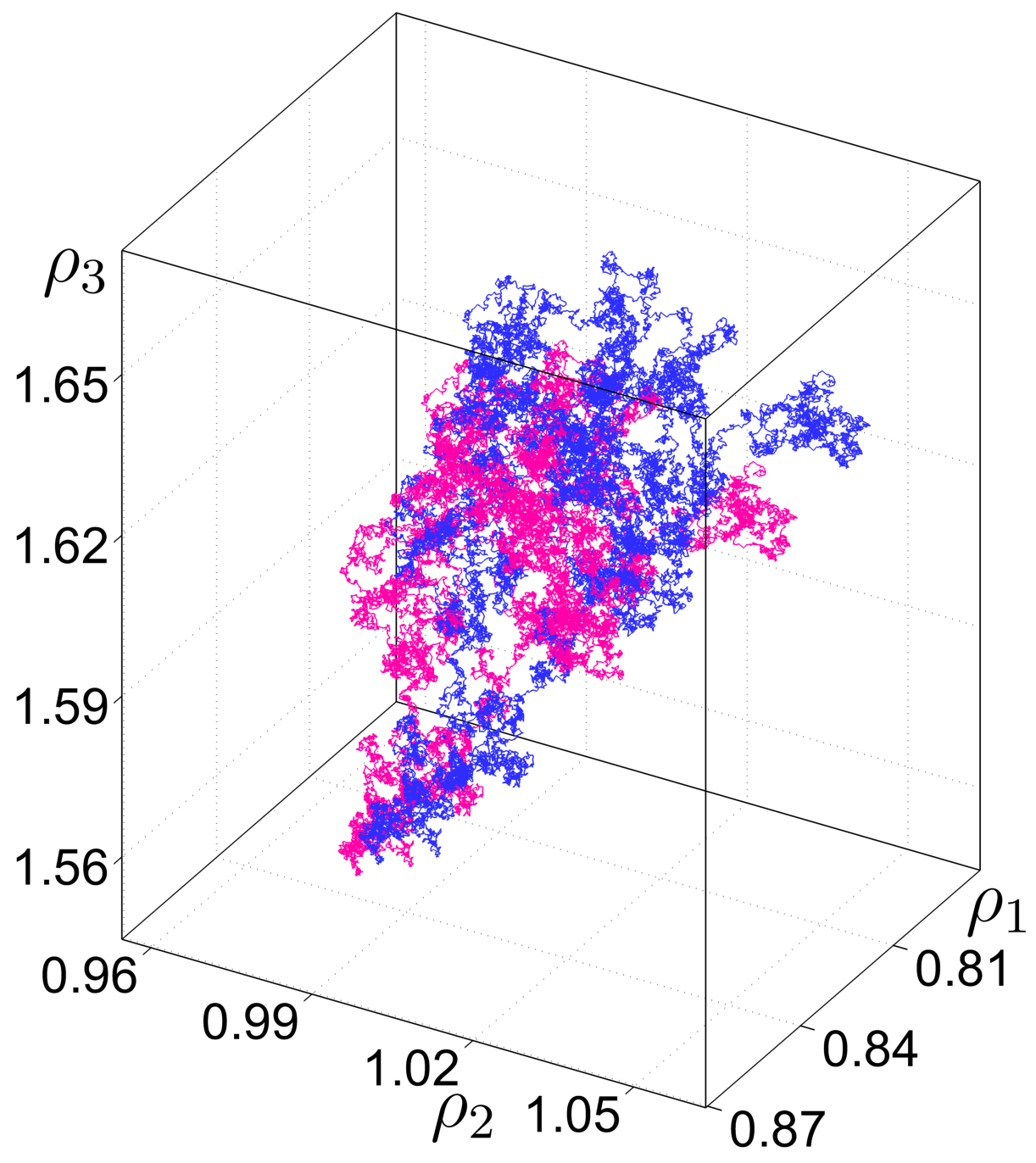}
\qquad
\includegraphics[width=65mm]{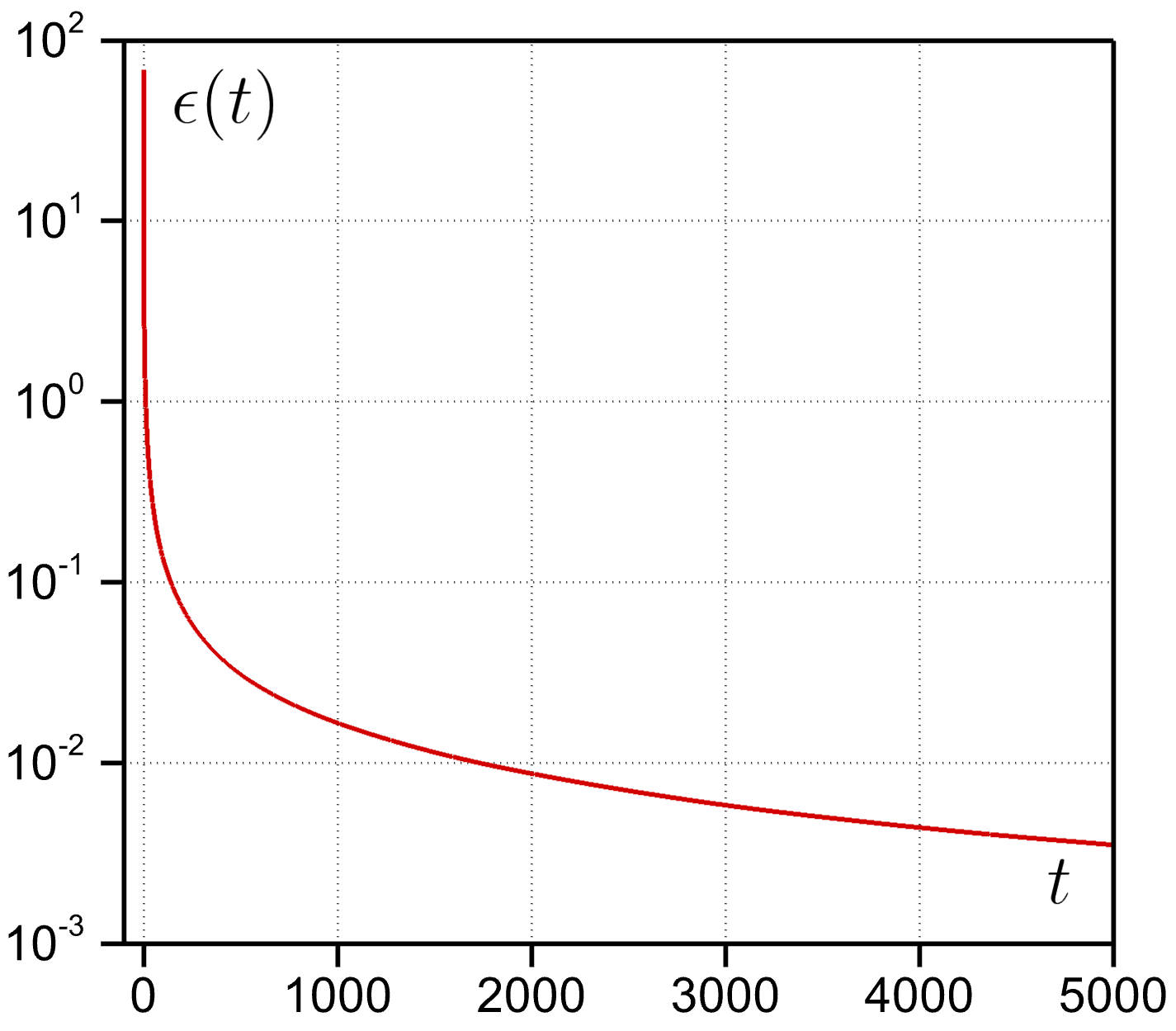}
\caption{\emph{On the left are plots of two internal times $\mathfrak{s}_1(\{\bar{\rho}\})$-(red curve) and  $\mathfrak{s}_2(\{\bar{\rho}\})$-(blue curve), which were obtained by calculating on the $\mathcal{A}_1$ manifold 
with initial conditions differing by $10^{-2}$. On the right is a plot of the Lyapunov exponent versus time. As can be clearly seen from the plot, the Lyapunov exponent very slowly tends to zero.}}
\label{Fig8}
\end{figure} 
Note that as the usual calculation time “$t$” increases, the internal time $\mathfrak{s}(\{\bar{\rho}\})$ in all considered cases fills the global three-dimensional space like the three-dimensional Hilbert curve \cite{Hilbert}, but with the 
only difference that now the filling occurs chaotically and inhomogeneously.

It is known, the Lyapunov exponent is an important indicator of the behavior of a dynamical system. In particular, when it is positive, i.e. $\epsilon(t)>0$, then we can talk about the chaotic behavior of the system of particles. As 
calculations on both manifolds show (see Fig.s 8 and 9), the Lyapunov exponent is positive. Moreover, for both cases under consideration, depending on the usual time “$t$”, both Lyapunov exponents very slowly tend to zero.
Note that such a fairly stable behavior of the Lyapunov exponent is very important for deriving Equations for the probabilistic distribution of the flow of geodesic trajectories in phase space (\ref{q2.0t9}). The latter, in turn, makes 
it possible to construct the entropy of a low-dimensional dynamical system (\ref {q3.K0wt}), estimate the measure of its complexity (\ref{q3.N0wt}) and, finally, estimate the degree of its “disequilibrium” (\ref{q5.N0wt}).
\begin{figure}
\quad
\includegraphics[width=65mm]{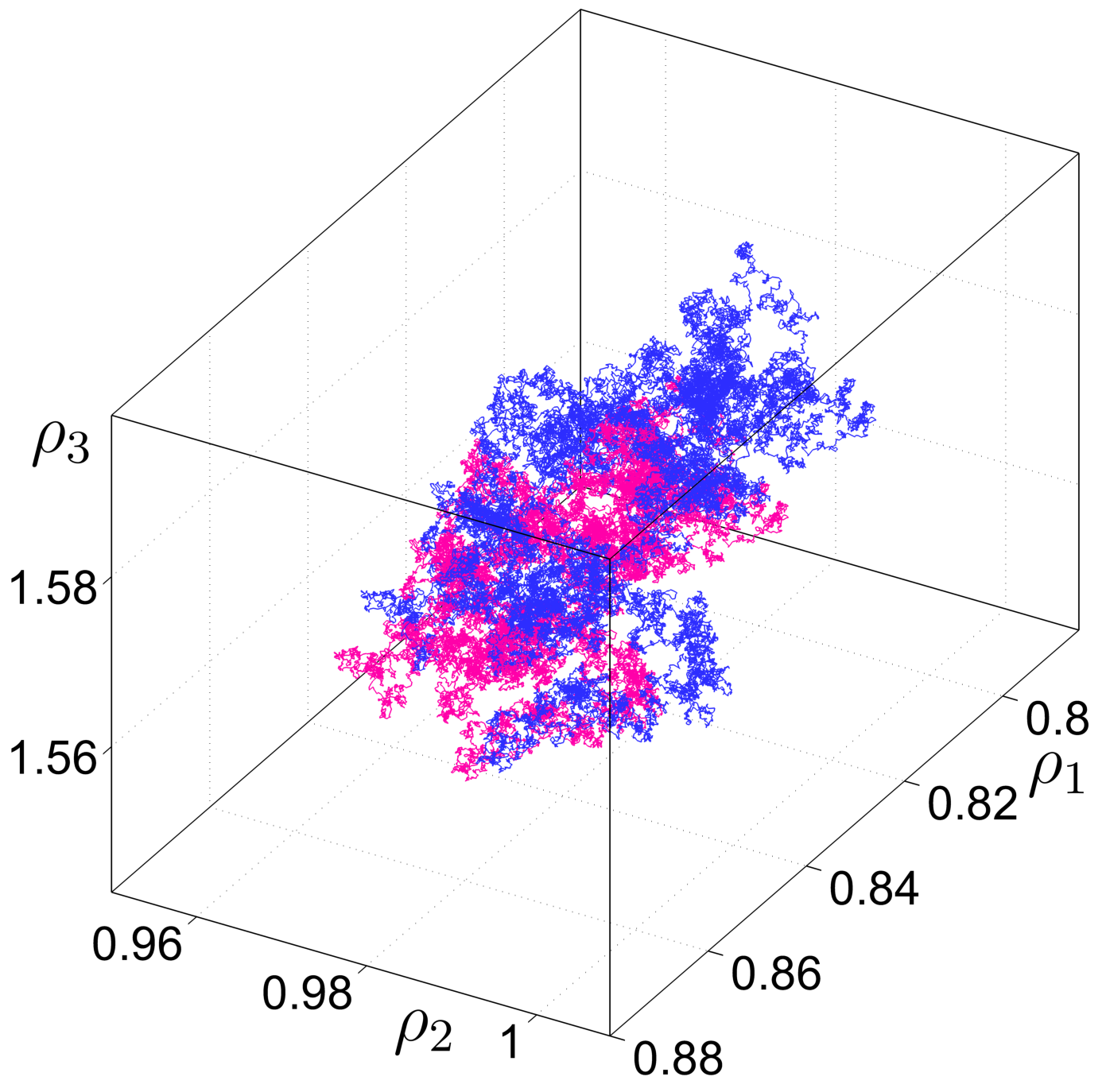}
\qquad
\includegraphics[width=65mm]{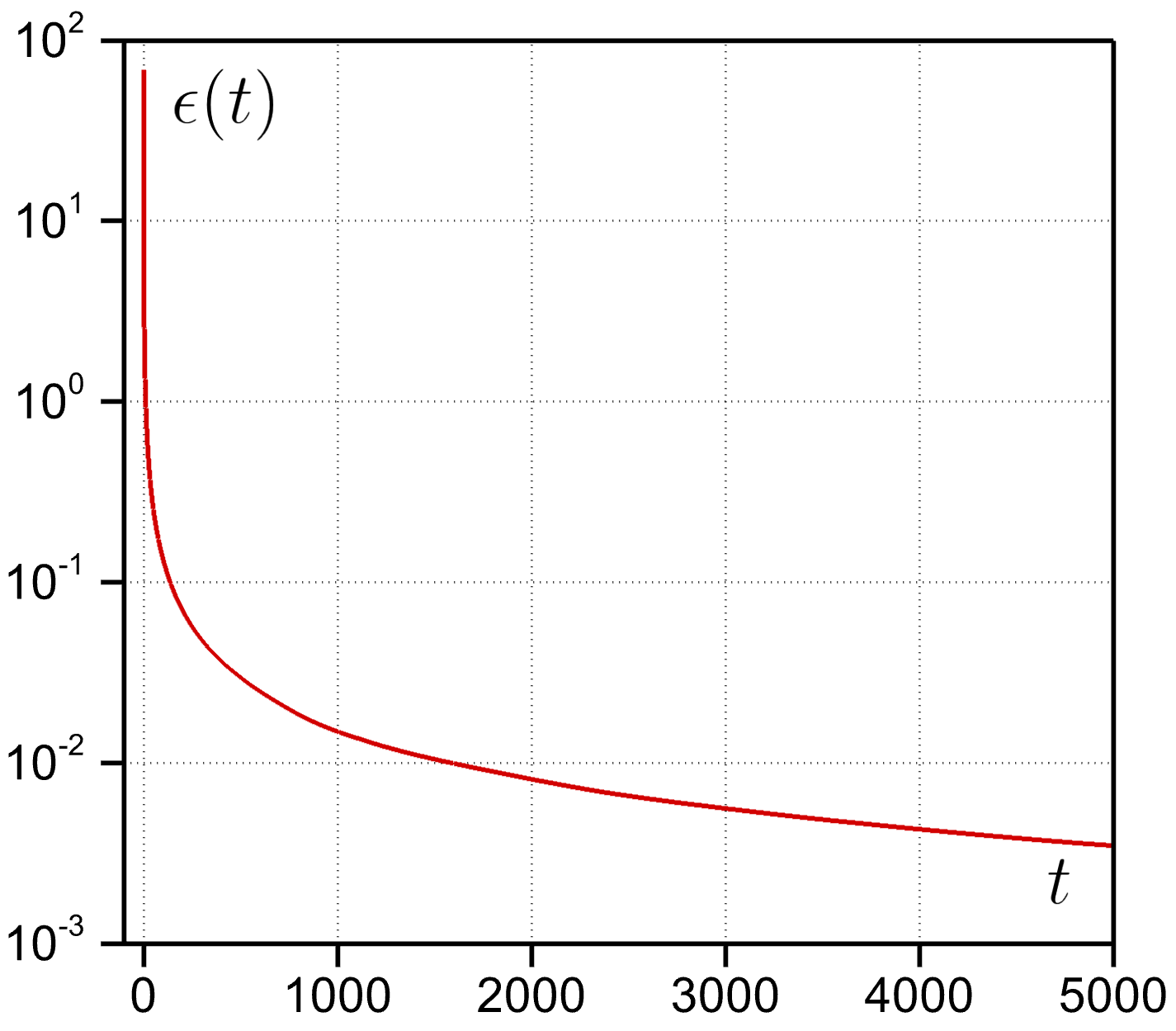}
\caption{\emph{On the left are plots of two internal times $\mathfrak{s}_1(\{\bar{\rho}\})$-(red curve) and  $\mathfrak{s}_2(\{\bar{\rho}\})$-(blue curve), which were obtained by calculating on the $\mathcal{B}_1$ manifold
 with initial conditions differing by $10^{-2}$. On the right is a plot of the Lyapunov exponent versus time.}}
\label{Fig9}
\end{figure} 

Important characteristics of the internal time of a restricted three-body problem are its macroscopic and microscopic structures in three-dimensional space. 

As calculations show,  the internal time for the selected initial data continuously fills a certain restricted three-dimensional region, which most likely indicates a certain degree of ergodicity of the dynamical system. In 
any case, to be sure of the ergodicity of a low-dimensional dynamical system, it is necessary to carry out additional detailed calculations and studies. It is very possible that for the system under consideration there 
will be areas of initial data where the ergodicity of the system will be partially or completely violated, for example, in the case of detection of periodic trajectories.  However, all this is the subject of new, more detailed
research.
\begin{figure}
\includegraphics[width=48mm]{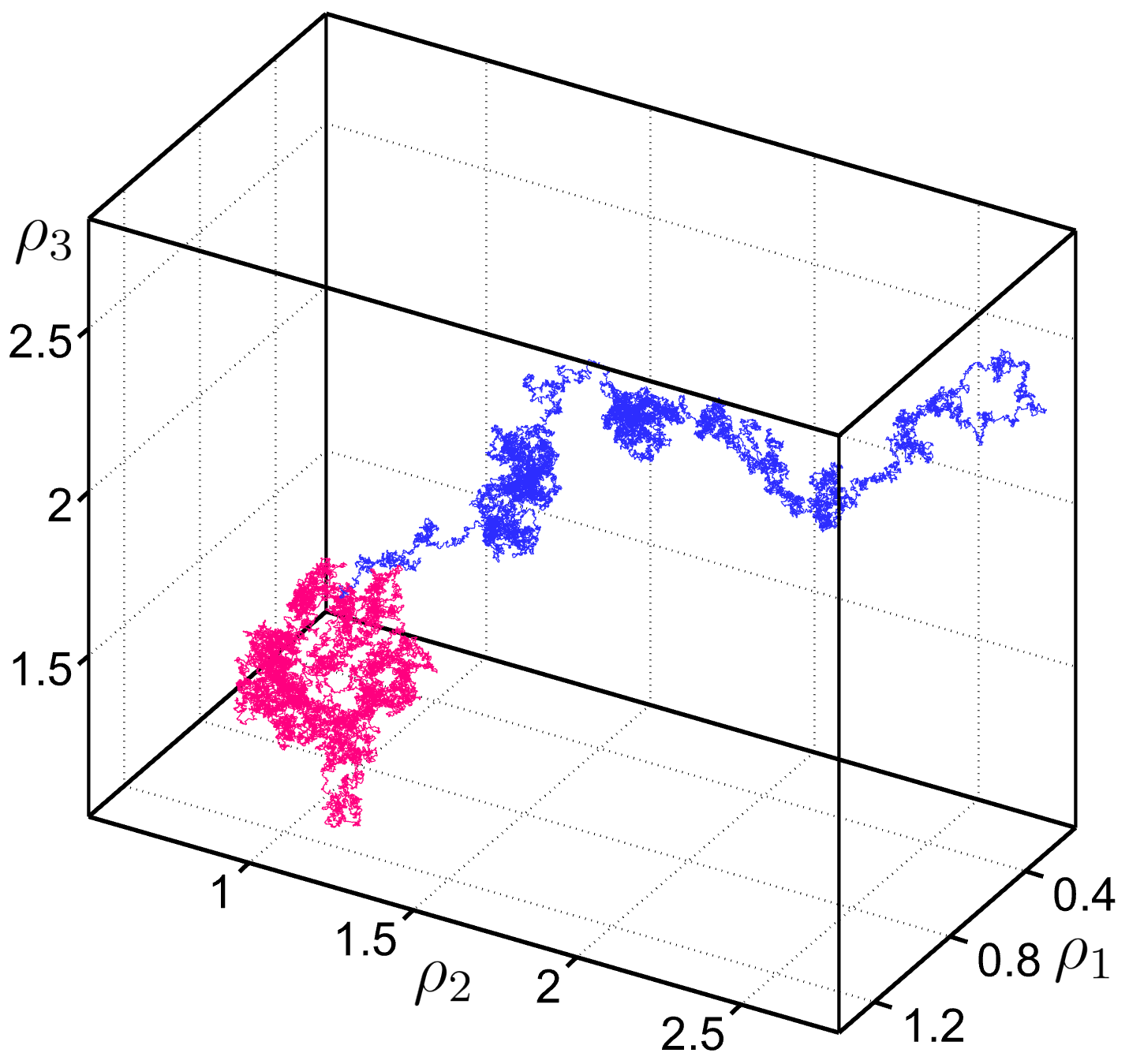}
\includegraphics[width=46mm]{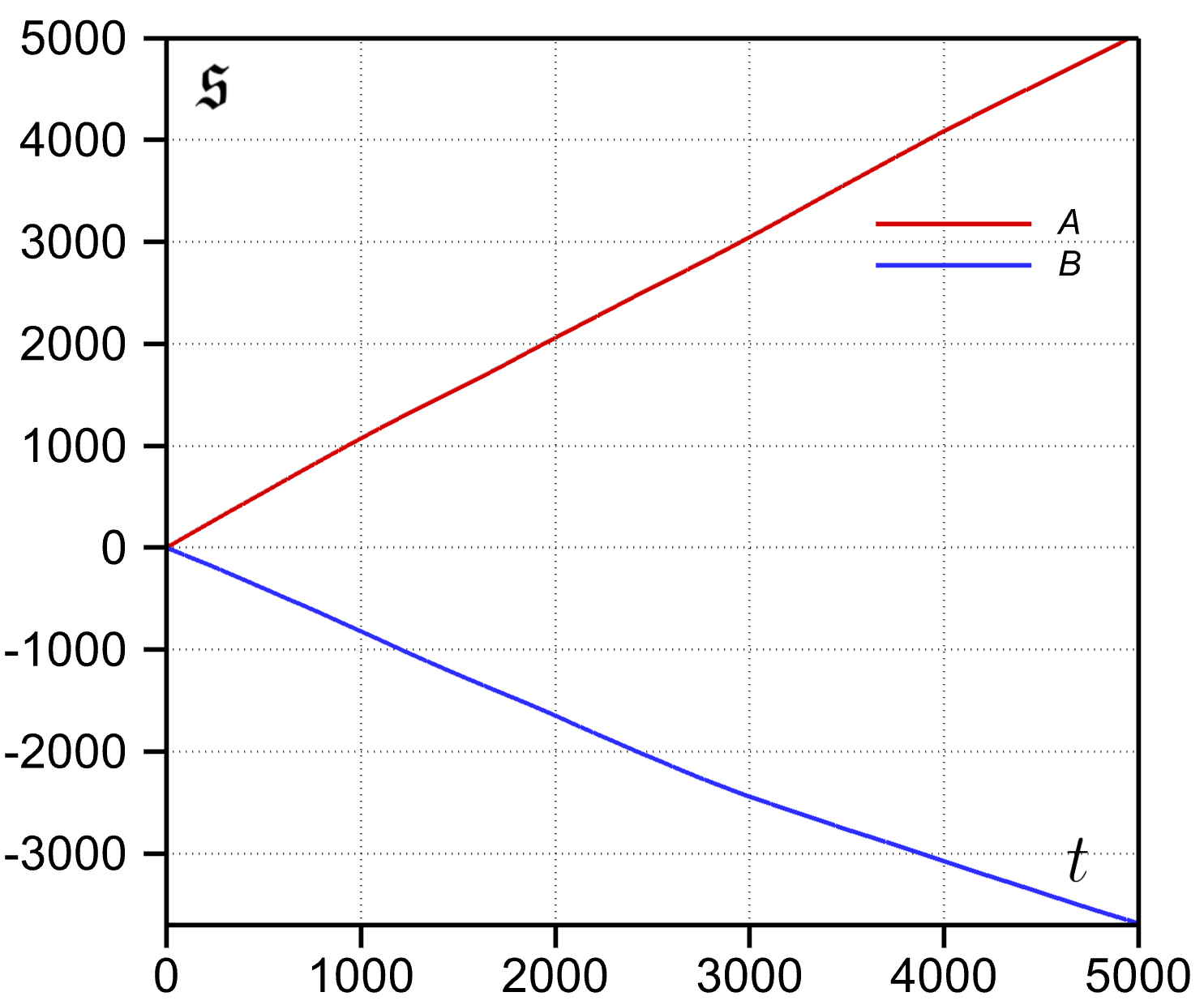}
\includegraphics[width=42mm]{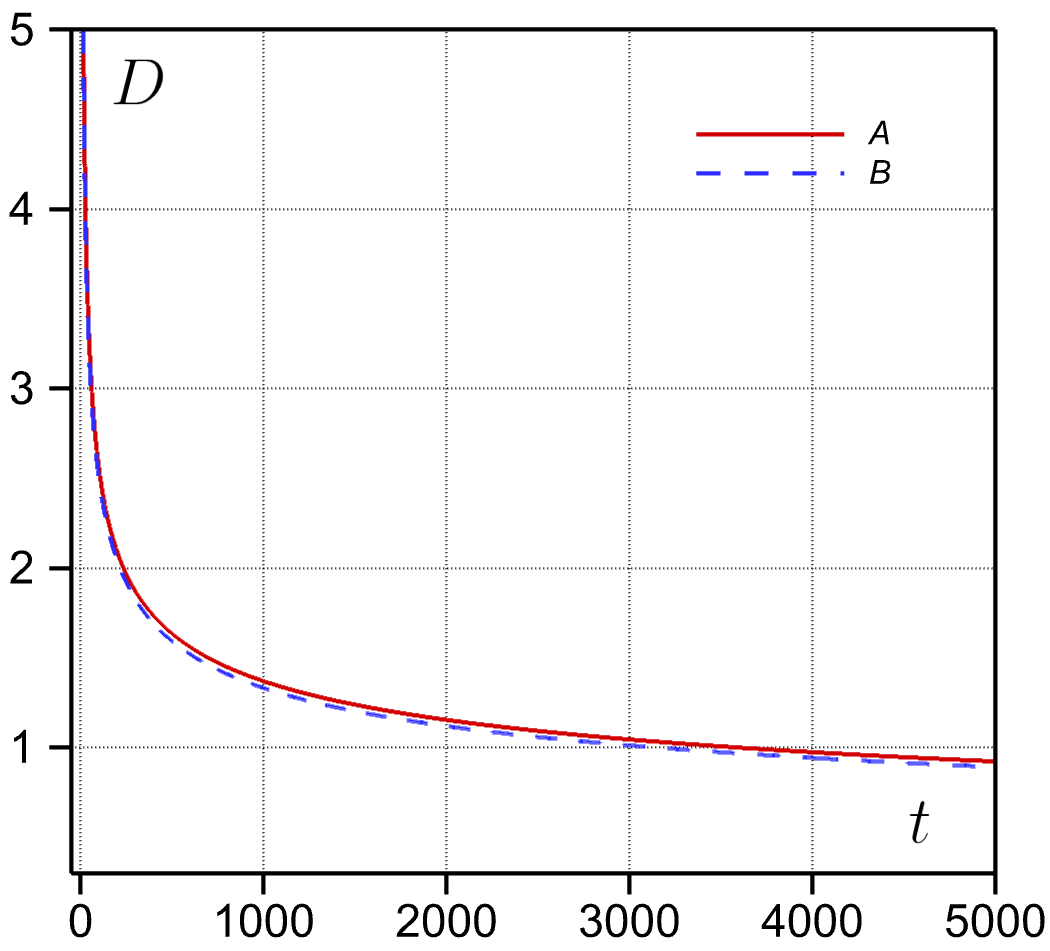}
\caption{\emph{On the left in the first figure, internal times $\mathfrak{s}_1(\{\bar{\rho}\})$-(red curve) and $\mathfrak{s}_2(\{\bar{\rho}\})$-(blue curve) are shown that were calculated on the manifolds' families 
$\mathcal{A}_1$ and $\mathcal{B}_1$ for the same initial data using the third line of {\bf Table 2}. The second plot from the left shows the internal time $\mathfrak{s}(t)$ depending the ordinary time ``$t$'' for the 
two marked families of manifolds. As can be seen from the graph, internal time can be either positive or negative. The third figure from the left shows the dimensionality of the structures formed by internal times in 
three-dimensional space.}}
\label{Fig10}
\end{figure} 
Now let us move on to the question, namely: does the random curve have a characteristic feature on large scales?

In particular, as can be seen from the plots in Fig.s  8 and 9, internal time at different scales is rather random self-similar three-dimensional curves, similar to stochastic  fractals, which are known to be non-deterministic \cite{Rakibur}.  
In this regard, a natural question arises: is there any universal characteristic of the three-dimensional random structure formed by internal time?

 As well-know,   the Hausdorff–Besicovitch exponent is used to determine the dimension of algebraic or geometric fractals, and therefore we cannot use it to calculate the dimension of structures such as a three-dimensional 
 random curve, which is internal time \cite{Falcone}. Recall that random fractals use stochastic rules; for example, L\'{e}vy flight,  self-avoidance walks, Brownian motion trajectories and Brownian tree, etc. The structure we 
 are studying is close to the concept of the trajectory of Brownian motion, which is a random fractal. Taking this into account, we propose a new criterion for determining the dimension of a stochastic three-dimensional fractal
 structure.

 {\bf Definition 4}. \emph{Let us call the dimension of a stochastic three-dimensional fractal structure the following expression:}
\begin{equation}
 {D}(t)=  {\ln\biggl\{\frac{1}{t}\Bigl|\int_0^{t}\mathfrak{s}(t')dt'\Bigr|\biggr\}}\Bigl/{\ln t}.
\label{qV.11}
\end{equation}
 
Using the data from the third row of {\bf Table 2}, we calculated the internal time on the families of manifolds $\mathcal{A}_1$ and  $\mathcal{B}_1$ (see Fig. 10).  As can be seen from the second figure, internal time 
as a function of ordinary time is a monotonic function in both cases considered, however, as numerical calculations show, non-monotonic dependencies are also possible, in the case of other initial data. In the third figure on the
left, the dimension of fractal structures filling the three-dimensional space of two internal times is calculated. It is shown that the dimension of both structures tends to its limiting asymptotic value $D=0.89$ at $t\geq5\cdot 10^3$.

In the end,  we note that, as preliminary calculations show, the asymptotic value of the dimension of internal time' structures in a number of areas of the initial data of the problem can change and even significantly. We will 
not touch on these problems, since they require more detailed study.

\section{C\lowercase{onclusion}}

We recently studied the general three-body problem, which, on the one hand, is quite well studied in classical mechanics and mathematics, and on the other hand, is a referance example of a low-dimensional dynamic 
system with complexity, a number of aspects of which still remain unexplored \cite{Gev0}. The general three-body problem is formulated in conformal Euclidean space, which is connected with the energy surface 
of the system of bodies and reflects all  its features. We formulated the problem in a conformal Euclidean-Riemannian space, which is associated with the energy surface of the system of bodies and reflects all its features. 
This formulation of the problem allows us to identify new hidden symmetries of the internal motion of a dynamic system and reduce its description instead of the known one - an eighth-order system to a sixth-order system 
(\ref{q1.01}). It is shown that a new parameter for chronologizing events of a dynamical system - internal time $\mathfrak{s}(\{\bar{\rho}\})$, which represents the trajectory of the center of mass of a system of bodies, 
in contrast to the usual time ``$t $'' has a number of unusual properties. In particular, as calculations show, for the limited three-body problem, the  parameter $\mathfrak{s}(\{\bar{\rho}\})$ is in the general case irreversible, 
multidimensional, inhomogeneous and oriented (has an arrow), which allows the system to choose the preferred asymptotic scattering subspace during the evolution  process (see diagram in Fig. 1 and Fig. 3). In other words,
a classical dynamic system consisting of three or more bodies always has a hidden, so-called internal time $\mathfrak{s}(\{\bar{\rho}\})\mapsto\mathfrak{s}(t)$, which has all the above properties.

The work examines in detail the mechanism  of the occurrence of randomness in the geodesic trajectory problem of a deterministic system. Recall  that this is due to the need to transition to a global reference system to carry 
out calculations, which is ensured by an underdetermined system of nonlinear algebraic equations (\ref{q1.04}). Within the framework of the representation of internal time, the Lyapunov time-depended exponent $\epsilon(t)$ 
(see (\ref{q2.G07})) was obtained, which determines the rate of divergence of close geodesic trajectories in the flow.  Assuming that the randomness in the system has a Gauss-Markovian character with the fluctuation power 
$\epsilon(t)$, stochastic differential Equations of Langevin type are written for the motion of the representing point with effective mass $\mu_0$. Using these SDEs, a second-order partial differential equation is obtained that 
describes the evolution of the probability density of the flow of geodetic trajectories of a low-dimensional dynamical system in the regime of chaotic motion.

An important achievement of the work is that if you use the representation of internal time to construct the corresponding quantum problem, then we can, without violating the well-known Arnold-Berry theorem \cite{Shust} in the 
limit $\hslash\to0$, move from the region of quantum motion to the region of classical chaotic Poincar\'{e} motion, and thereby solve the the problem of quantum-classical reconciliation.  We emphasize that it is precisely this f
ormulation of the problem that can lead to chaos in the behavior of the wave function of a deterministic quantum system. Recall  that the criterion for the emergence of quantum chaos is the inequality $I^0(t)>\hslash^3$, 
where  $I^0(t)$ is the volume of the region of classical chaos in phase space.

To test the developed representation using numerical methods, we considered the problem of three bodies interacting with each other through Morse potentials. For definiteness, we carried out calculations on two complete 
members of the families of manifolds  $\mathcal{A}_1$ and $\mathcal{B}_1$.

Based on the fact that under certain initial conditions a limited three-body problem arises, we conducted a detailed study of the properties and structure of three-dimensional internal time, which describes, on the other hand, 
the trajectory of the effective mass $\mu_0$ in global space $\{\bar{\rho}\}\in\mathbb{R}^3$. As numerical experiments have shown, even in this particular case, internal time $\mathfrak{s}(t)$, in depending of usual time
 ``$t$'' fills three-dimensional space chaotically and inhomogeneously. The work studies in detail the issue of the dimension of stochastic fractals generated by a dynamical system and defines a new criterion for it (see {\bf Definition 4}). 
 As shown, for large times $t \geq 5\cdot10^3$ the dimension of three-dimensional structures formed by internal time in the case of both manifolds is approximately equal $D=0.89$.

We expect that the study of the scattering problem in a three-body system within the framework of a new representation \cite{Gev0} will reveal new interesting features of internal time $\mathfrak{s}(t)$, in particular, it will clearly 
demonstrate the actions of the arrow of time when choosing a specific type of elementary atomic-molecular reaction or a certain scattering subspace (see Fig. 1) .
 
In conclusion, we note that the developed approach together with the created algorithm for the first time provides real opportunities with a given accuracy for the numerical study of a multichannel atomic-molecular process within 
the framework of classical mechanics, taking into account the difficulties inherent in the integration of rigid systems of ordinary differential equations.


\end{document}